%% file: main.tex
\documentclass[10pt]{IEEEtran}
\usepackage[left=1.62cm,right=1.62cm,top=1.9cm]{geometry}
\usepackage{cite}
\usepackage{amsmath,amssymb,amsfonts}
\usepackage{algorithm}
\usepackage{algpseudocode}
\usepackage{graphicx}
\usepackage{textcomp}
\usepackage{xcolor}
\usepackage{multirow}
\usepackage[normalem]{ulem}
\usepackage[T1]{fontenc}
\usepackage{svg}

\def\BibTeX{{\rm B\kern-.05em{\sc i\kern-.025em b}\kern-.08em
    T\kern-.1667em\lower.7ex\hbox{E}\kern-.125emX}}
\usepackage{url}

\begin{document}

\title{Energy-Efficient Computation with DVFS using Deep Reinforcement Learning for Multi-Task Systems in Edge Computing}

\author{
Xinyi Li\IEEEauthorrefmark{2}\IEEEauthorrefmark{1}, Ti Zhou\IEEEauthorrefmark{3}\IEEEauthorrefmark{1}, Haoyu Wang\IEEEauthorrefmark{4} and Man Lin\IEEEauthorrefmark{5}
\thanks{Department of Computer Science,
St. Francis Xavier University,
Canada. 
Email: 
\IEEEauthorrefmark{2}x2021gim@stfx.ca,
\IEEEauthorrefmark{3}tizhou1@cs.stonybrook.edu,
\IEEEauthorrefmark{4}x2020fcw@stfx.ca,
\IEEEauthorrefmark{5}mlin@stfx.ca.}
\thanks{\IEEEauthorrefmark{1}These authors contributed equally to this work.}
\thanks{The research is supported by the Natural Sciences and Engineering Research Council of Canada. \\ A preprint appears at https://arxiv.org/abs/2409.19434}
}

\maketitle

\begin{abstract}
Finding an optimal energy-efficient policy that is adaptable to underlying edge devices while meeting deadlines for tasks has always been challenging. This research studies generalized systems with multi-task, multi-deadline scenarios with reinforcement learning-based DVFS for energy saving for periodic soft real-time applications on edge devices.  This work addresses the limitation of previous work that models a periodic system as a single task and single-deadline scenario, which is too simplified to cope with complex situations.  The method encodes time series data in the Linux kernel into information that is easy to interpret for reinforcement learning, allowing the system to generate DVFS policies to adapt system patterns based on the general workload. For encoding, we present two different methods for comparison. Both methods use only one performance counter: system utilization, and the kernel only needs minimal information from the userspace. Our method is implemented on Jetson Nano Board (2GB) and is tested with three fixed multitask workloads, which are three, five, and eight tasks in the workload, respectively. For randomness and generalization, we also designed a random workload generator to build different multitask workloads to test. Based on the test results, our method could save 3\%-10\% power compared to Linux built-in governors.

\end{abstract}


\input{Sections/introduction_v2}

\input{Sections/proposed_method}

\input{Sections/experiment}

\input{Sections/related_work}
\input{Sections/conclusion}

\bibliographystyle{IEEEtran}
\bibliography{reference}

%temp
% \begin{IEEEbiographynophoto}{Jane Smith}
% Jane Smith is a professor at ABC University. Her research focuses on energy-efficient computing and green data centers.
% \end{IEEEbiographynophoto}

% \begin{IEEEbiography}{\includegraphics[width=1in,height=1.25in,clip,keepaspectratio]{john_doe.jpg}}{John Doe}
% John Doe received his Ph.D. in Computer Science from XYZ University. His research interests include sustainable computing and distributed systems.
% \end{IEEEbiography}

\begin{IEEEbiographynophoto}{Xinyi Li} 
Xinyi Li received her Bachelor's degree in Computer Science from St. Francis Xavier University, Antigonish, Canada. She is currently pursuing a Master's degree at Dalhousie University, Halifax, Canada. Her research focuses on cybersecurity and machine learning.
\end{IEEEbiographynophoto}

\begin{IEEEbiographynophoto}{Ti Zhou}
Ti Zhou received his Bachelor’s and Master’s degrees in Computer Science from St. Francis Xavier University, Antigonish, Canada. He is currently pursuing a Ph.D. in Computer Science at Stony Brook University, USA. He has a broad interest in computer systems, with a current focus on verification of distributed systems and energy efficiency.
\end{IEEEbiographynophoto}

\begin{IEEEbiographynophoto}{Haoyu Wang}
Haoyu Wang received the B.S. degree in Computer Science through a joint undergraduate program between Changzhou University, Changzhou, China and St. Francis Xavier University, Antigonish, Canada in 2023. He is currently pursuing an M.S. degree in Computer Science with the Faculty of Computer Science, Dalhousie University, Canada. His research interests include Linux kernel optimization, dynamic voltage and frequency scaling (DVFS), and energy performance trade-off in embedded systems.
\end{IEEEbiographynophoto}

\begin{IEEEbiographynophoto}{Man Lin}
Man Lin received her B.E. degree in Computer Science and Technology from Tsinghua University, China, in 1994. She received the Lic. and Ph.D. degrees from the Department of Computer Science and Information at Linkoping University, Sweden, in 1997 and 2000, respectively. She is currently a Professor of Computer Science at St. Francis Xavier University, Canada. Her research interests include real-time and cyber-physical system design and analysis, scheduling, power-aware computing, and optimization algorithms. Her research is supported by the National Science and Engineering Research Council of Canada (NSERC). 
\end{IEEEbiographynophoto}

%\begin{IEEEbiography}[{\includegraphics[width=1in,height=1.25in,clip,keepaspectratio]{Figure/Haoyu.jpg}}]{Haoyu Wang}
%Haoyu Wang received the B.S. degree in Computer Science through a joint undergraduate program between Changzhou University, Changzhou, China and St. Francis Xavier University, Antigonish, Canada in 2023. He is currently pursuing the M.S. degree in Computer Science with the Faculty of Computer Science, Dalhousie University, Halifax, Canada. His research interests include Linux kernel optimization, dynamic voltage and frequency scaling (DVFS), and energy-performance tradeoff in embedded systems.
%\end{IEEEbiography}

\end{document}

%% file: Sections/introduction_v2.tex
\section{\textcolor{black}{Introduction}}
\textcolor{black}{Soft real-time systems play a critical role in various domains by enabling the processing of periodic tasks that are time-constrained but not strict. They have many applications in multiple fields.} 
Here are some examples: an optimization control software for automobiles \cite{minaeva2020control} that involves periodic sensing, computing, and driving tasks with real-time requirements; a smart farm system where most communication is periodic \cite{nobrega2019iot}; and a system that periodically monitors soil health \cite{ramson2021self}.

For these systems, performance and energy are two indicators that interest researchers. \textcolor{black}{However, a system that can achieve high performance and save energy at the same time is almost impossible to realize, so achieving a balance between performance and energy consumption in such a soft real-time system is a key challenge.}
A previous work \cite{zhou2023cpu} modelled them as a single task with a single deadline system and proposed a model-free method to save energy for the system while ensuring performance.
However, this modelling method has its limitations because of oversimplification.
In real-world applications, there is usually more than one task in one period. Some of these tasks could contain dependencies that need to be executed sequentially, such as getting and processing data \cite{lee2022probabilistically} \cite{teslyuk2020optimal} \cite{manogaran2020isof}. Some tasks can be executed in parallel, such as the system using multiple sensors which can acquire data simultaneously \cite{salman2021systematic} \cite{liu2020continuous} \cite{yuan2021design}. 
\textcolor{black}{This shows another challenge in managing the complexity of soft real-time periodic systems.}

The previous work \cite{zhou2023cpu} could not adequately model the complex situation with only a single task and a single deadline. In this work, we study systems that have multiple tasks in one period, and each of these tasks has a deadline. 

We consider multi-tasks in a task set that runs periodically. 
Each task $task_{i}$ has its start timestamp $start_i$, execution time $exet_{i}$, finish timestamp $finish_{i}$ and deadline $ddl_{i}$. For simplicity, we assume that each task is independent of one another, and their start time depends only on the predefined setting time. During each period, the task set is executed once for a total deadline time of $T$, which is the deadline for the last task. Other deadlines cannot exceed $T$.

\begin{figure*}[h]
  \centering
  \includegraphics[width=\linewidth]{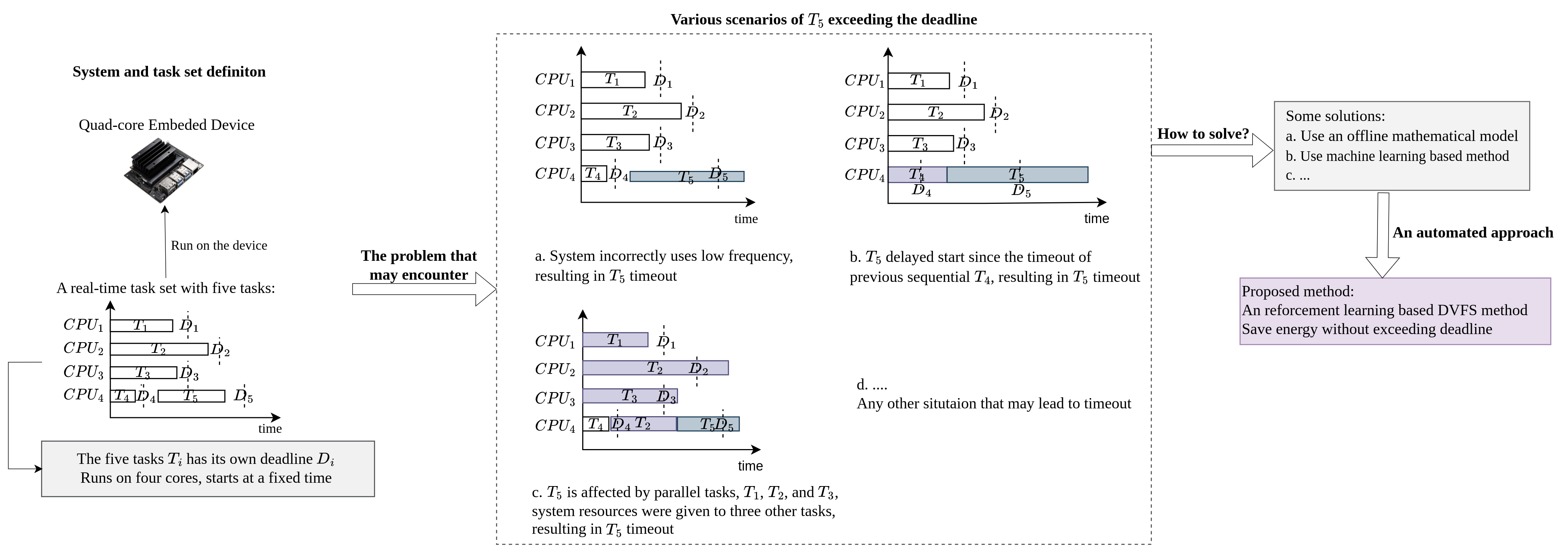}
  \caption{Deadline Missing Example: A Multi-Task set with Multi-Deadline on Multi-Core
  %System definition and method to solve.
  }
  \label{fig:definition}
\end{figure*}

Fig.~\ref{fig:definition} gives a detailed example of the type of systems studied in this paper. In this example, a real-time task set consists of five tasks that run on a device with four cores. This series of tasks runs periodically on the system. Our aim is to find an adaptable policy that can save as much system energy as possible without causing tasks to miss their deadlines on the multi-core system. The only reason a deadline is missing on a single-tasking system is that the system does not run with a high enough frequency.
However, various situations may cause a timeout during a run on multicore systems, as shown in Fig.~\ref{fig:definition}.  On a multi-core system, a task may miss the deadline because appropriate frequencies are not used, the same as on single-tasking systems. A previous task on the same core or tasks on other cores that have migrated to this core may also cause the deadline to be missed.
Thus, in a multi-task system, the causes for task timeouts become much more complex. How can the frequencies be managed without causing tasks to miss their deadlines and save as much energy as possible? Possible solutions are to construct offline mathematical models 
%\cite{chang2020real} %\cite{xu2024energy} 
\cite{wu2023energy} \cite{houssam2020hpc} \cite{chen2020scheduling} or to solve the problem automatically with machine learning-based methods \cite{hoffmann2021online} \cite{zhan2019power} \cite{li2024deep} \cite{chen2021energy} \cite{fettes2018dynamic}. \cite{panda2022energy} \cite{zhou2021deadline} \cite{zhou2023cpu}.

To solve this problem, we propose to combine Reinforcement Learning (RL) and Dynamic Voltage and Frequency Scaling (DVFS) with an encoding of system state to capture the deadline of multi-tasks and multi-core system status and make predictions. DVFS is an energy-saving method which adapts voltage and frequency dynamically to satisfy the system's needs. Linux has some built-in DVFS governors \cite{zhou2023profiling}, such as \textit{Ondemand} and \textit{Conservative}. They make predictions based on the utilization in the previous period, that is, they assume the next period is the same as the previous one. Also, they adjust the frequency based on the max load of the system. Our method uses Double Deep Q Learning (DDQN) \cite{van2016deep} to train a DVFS policy since every execution of the task set which reaches $T$ time could be seen as an episode. This task set will be executed each period which indicates that the events that happened in each period are almost the same. The governor could predict the events and system states once a good policy is driven.

\begin{figure}[h]
  \centering
  \includegraphics[width=\linewidth]{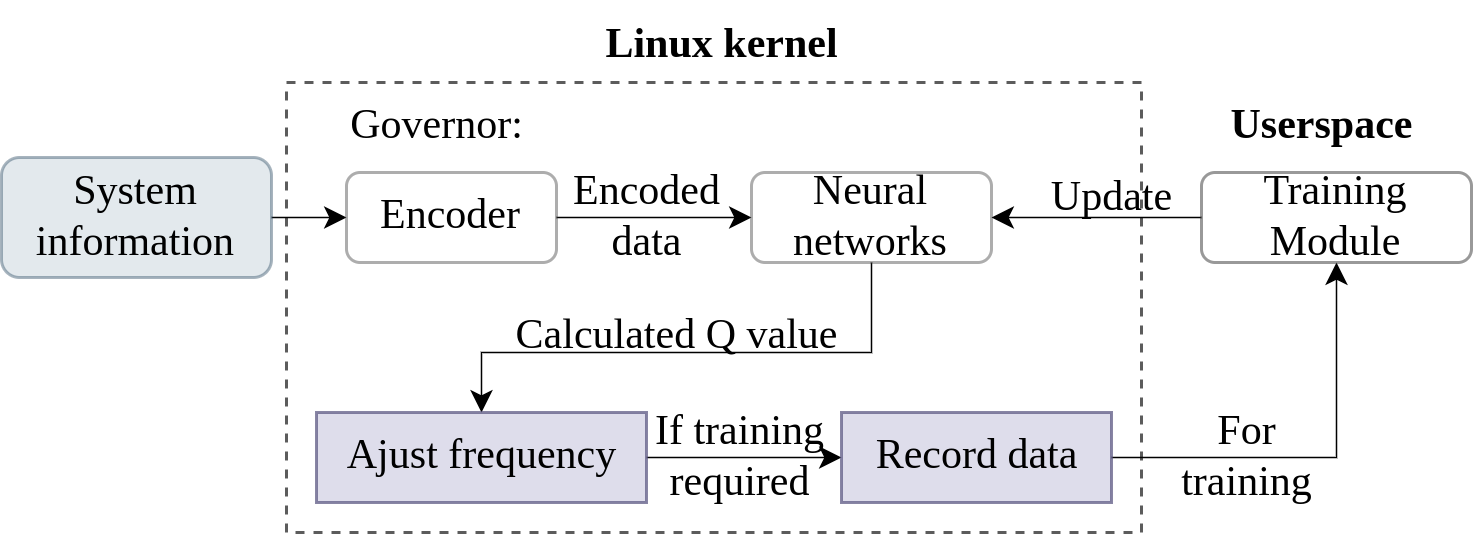}
  \caption{An overview of our proposed method.}
  \label{fig:overview}
\end{figure}

An overview of our method is shown in Fig.~\ref{fig:overview}. In the Linux kernel, we insert an encoder and neural network. The encoder is responsible for encoding the system information from the last period and the observations from this period. The encoded data will be sent to networks with frequency as action to calculate the corresponding Q value. Based on the max Q value, the governor chooses the corresponding frequency for the next period. If training is required, all the data will be recorded and sent to userspace to update the networks. The new network will be sent to the kernel for testing. After around 400 trainings, the governor could get a good policy. 

Our proposed method is more accurate than traditional Linux built-in governors based on the following reasons. Our method uses the temporal series information in the kernel including average and max utilization. While the Linux built-in governors only use max load from the last period which means they have little information and do not have an overall judgement of the workload and system status.

Compared with the previous method for a single task and single deadline situation \cite{zhou2023cpu}, our new method is not a pure model-free strategy as the task number needs to be passed to the kernel. Besides this information, the method could encode the system parameters to useful information which be learnt by RL. 

\textcolor{black}{Our contributions are:}
\begin{itemize}
    \item \textcolor{black}{To the best of our knowledge, we are the first to propose a DVFS method  implemented at the Linux kernel level that can adapt the frequency based on multiple workloads and overall system status.}
    %overall workload and system status.}
    \item \textcolor{black}{Our proposed method targets multitasking with multiple deadlines in a soft real-time system which can save 3\% - 10\% power compared with a Linux built-in governor.}
    \item \textcolor{black}{The method targets small embedded devices with low number of frequency/voltage (F/V) levels and is implemented in a real device, Jetson Nano Board 2GB. However, the method is general and can be applied to other devices that have more F/V levels.}
\end{itemize}

%% file: Sections/proposed_method.tex
\section{Proposed Method}
\begin{figure}[h]
  \centering
  \includegraphics[width=\linewidth]{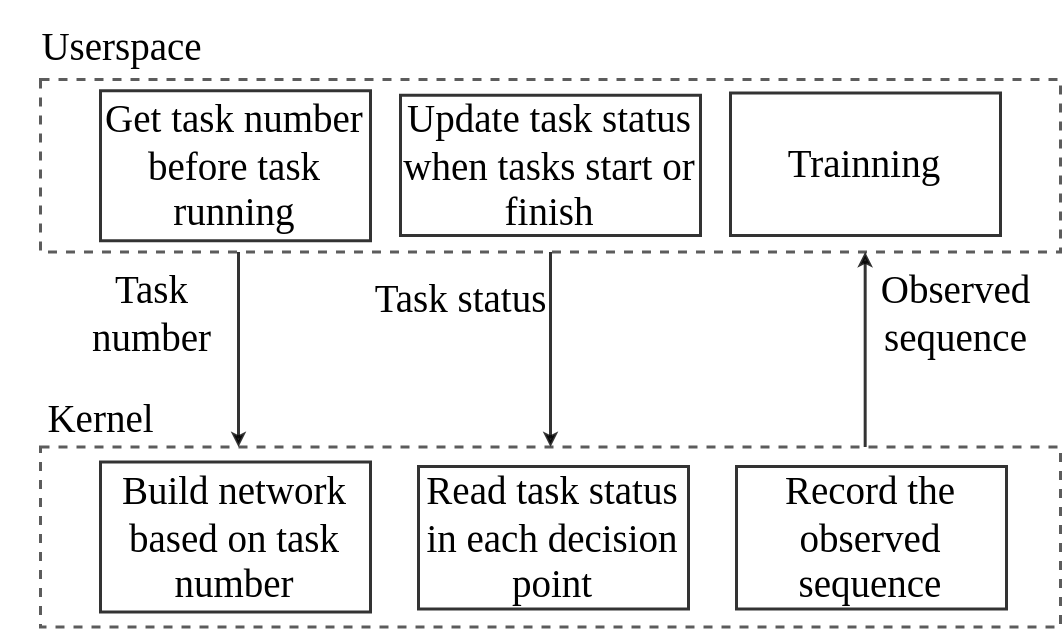}
  \caption{Interaction between userspace and kernel.}
  \label{fig:interaction}
\end{figure}

In this section, each part of our proposed method will be demonstrated in detail.
Similar to \cite{zhou2023cpu}, the structure of the method includes modules in both the kernel and the userspace. The Linux kernel consists of two parts for the driven policy to make decisions, one is the temporal encoder for building the environment state, and the other is the neural network for calculating the corresponding Q value. The training module is in userspace, using the reinforcement learning method, Double Deep Q Network. Fig.~\ref{fig:interaction} shows the information that is delivered between userspace and kernel. \textcolor{black}{The data that needs to be obtained by the kernel is the task number and status of each task. The kernel will record the observed sequence of the system when running the task and then pass them to the userspace for training.}

\subsection{Temperal Encoder}
\label{ch:encoding}
Each time the governor makes the decision, it could get information from the system and userspace. Based on these parameters, the governor could make a decision to adjust the frequency. The parameters observed from the system are CPU utilization, the runtime of the CPU so far, and the total time of it so far. The CPU total time consists of CPU runtime and CPU idle time. The parameters get from userspace is the label for each task, which marks each task's three possible execution statuses: has not started (0), is in progress (1) or has completed (2). This information consists of the time series that the governor observed. 
We have two encoding methods for the time series. The first encoder is for networks which have one layer Gated Recurrent Unit (GRU) and a two-layer Multilayer Perceptron (MLP). The output of GRU will be sent to MLP for processing. 
\textcolor{black}{This is the final encoder in our proposed method and we use PRO. for abbreviation. For comparison, we have another encoder which consists of only a two-layer MLP. We use PROM. as an abbreviation for the method using this encoder.}

\subsubsection{Temporal Encoder (TEGM) for Network using GRU and MLP}
The time series observed at time $t$ is encoded as five components, $s_{t} = (p1_{t}, p2_{t}, i_{t}, u_{t}, c_{t})$.
$p1_{t}$ \textcolor{black}{captures in what frequency and load-interval that the system is currently active and the corresponding time spent. It is represented as an array based on the period from the last governor's decision-making time.  For each frequency and each max-load-interval, the corresponding array element represents the active time in this combination. } For example, if we have two frequencies to select, high frequency and low frequency, and the max load is divided into two \textcolor{black}{intervals, one is between [0, 60] and the other is [60, 100]}, then the array will have four elements. The element is the period time if the current system runs with the corresponding frequency and load, otherwise it will be 0.
\textcolor{black}{This element avoids the drawbacks of the traditional governor, which selects the frequency based only on the load. It segments the load and is related to the frequency so that, at the end of the task, even with low system load, the frequency needs to be increased to avoid timeouts, thus avoiding situation (a) in Fig.~\ref{fig:definition}.}
%not clear

$p2_{t}$ is a multi-dimensional array denoting the start and finish time stamp for each task. One sub-array consists of these two pieces of information for one task. The time stamps are all timed from the start of the first task.
\textcolor{black}{The aim of designing this element is to let the system better capture the task pattern in each period since the predefined start time of each task is hidden for the kernel. Suppose there is a period in which one of the tasks starts at a different time. In that case, the system can potentially surmise that the task could be affected by other tasks and thus have the possibility of a timeout, as shown in situations (b) and (c) in Fig.~\ref{fig:definition}.}
%not clear

These two $p_t$'s comprise the information about multitasking. $p1_{t}$ is designed to obtain the influence on system frequency and load of all tasks currently being performed. Thus, the system could implicitly infer from $p1_{t}$ about the number of tasks currently being executed. $p2_{t}$ is to give the system an overall control over the execution phase of the task set. Based on the start time and end time of each task, the system can get the tasks that have been completed, those that are in the process of execution, and those that have not yet started.

\textcolor{black} {$i_{t}$ is an instance observation, which is a two-element array representing the maximum load and average load during the current period. This information can help the system understand the load status of multitasks currently being executed on the multi-core system.} 
%Due to the multi-core architecture and multi-tasking, this information can likewise help the system to surmise the number of tasks that are currently being executed. More importantly, in order to address the problem of uneven load distribution, it allows the system to develop a policy to tone down the frequency when the average load is high.

$u_{t}$ is the average \textcolor{black}{utilized run} time of the entire system accumulated so far. It is obtained by multiplying the time of each time period by the average load over that time period. \textcolor{black}{$u_t$ represents the overall average utilization compared to the total time consumed, $c_{t}$, described below.}

$c_{t}$ is the \textcolor{black}{consumed} time from the beginning of the entire task set. Combining this information with $p2_{t}$ and $u_{t}$, the system could get control of the execution status of the whole task set.

\begin{algorithm} 
    \caption{Temporal Encoder (\textcolor{black}{TEGM}) for network with GRU and MLP} 
    \label{alg:encoder} 
    \begin{algorithmic}[1]
    \State \textbf{INPUT:} The label $label_i$ sent from userspace for each task, the system observation of the last period $(freq, util_{max}, util_{avg})$, time consumption of last period $t$, \textcolor{black}{$start\_label_i$ and $end\_label_i$.}
    \State $s_{t-1} = (p1_{t-1}, p2_{t-1}, i_{t-1}, u_{t-1}, c_{t-1})$
    \State \textbf{OUTPUT:} $s_t$
    \State $p1_t[freq][load] = t$
    \For{each label $label_i$}
        \If {$label_i$ is 1 and $start\_label_i$ is 0}
            \State $p2_t[i][start] = c_{t-1}$
        \ElsIf{$label_i$ is 2 and $end\_label_i$ is 0}
            \State $p2_t[i][end] = c_{t-1}$
        \EndIf
    \EndFor
    \State $i_t = (util_{max}, util_{avg})$
    \State $u_t = u_{t-1} + t \times util_{avg}$
    \State $c_t = c_{t-1} + t$
    \State $s_{t} = (p1_{t}, p2_{t}, i_{t}, u_{t}, c_{t})$
    \end{algorithmic}
\end{algorithm}

Overall, the difficulty of multitasking compared to the single task situation is the handling of the information of multiple tasks. Here we use $p_t$ to solve this problem, which is separated into two parts. $p1_t$ is used to get information about the current system, and $p2_t$ is used to get information about the execution of each task so far. $u_{t}$ is to get the current load of the system. Both $i_t$ and $c_t$ are designed to get the execution status of the whole task set. 
\textcolor{black}{The main difference between the encoding of single-task and multitask is $p_t$. In single-task system, $p_t$ only has one part which is $p1_{t}$, as there is only one task in the system, and thus the task status is the system status. In the multitask system, the status of each task should be separately recorded in $p_t$, so $p1_{t}$ is used to record the system's frequency and load, and $p2_{t}$ is responsible for recording the task status. Other encoding parts are almost identical to those for single-task systems, except that the multitask system has no quantization. The reason for this will be explained in Section~\ref{sec:inferencemodule}.}

\begin{figure}[h]
  \centering
  \includegraphics[width=\linewidth]{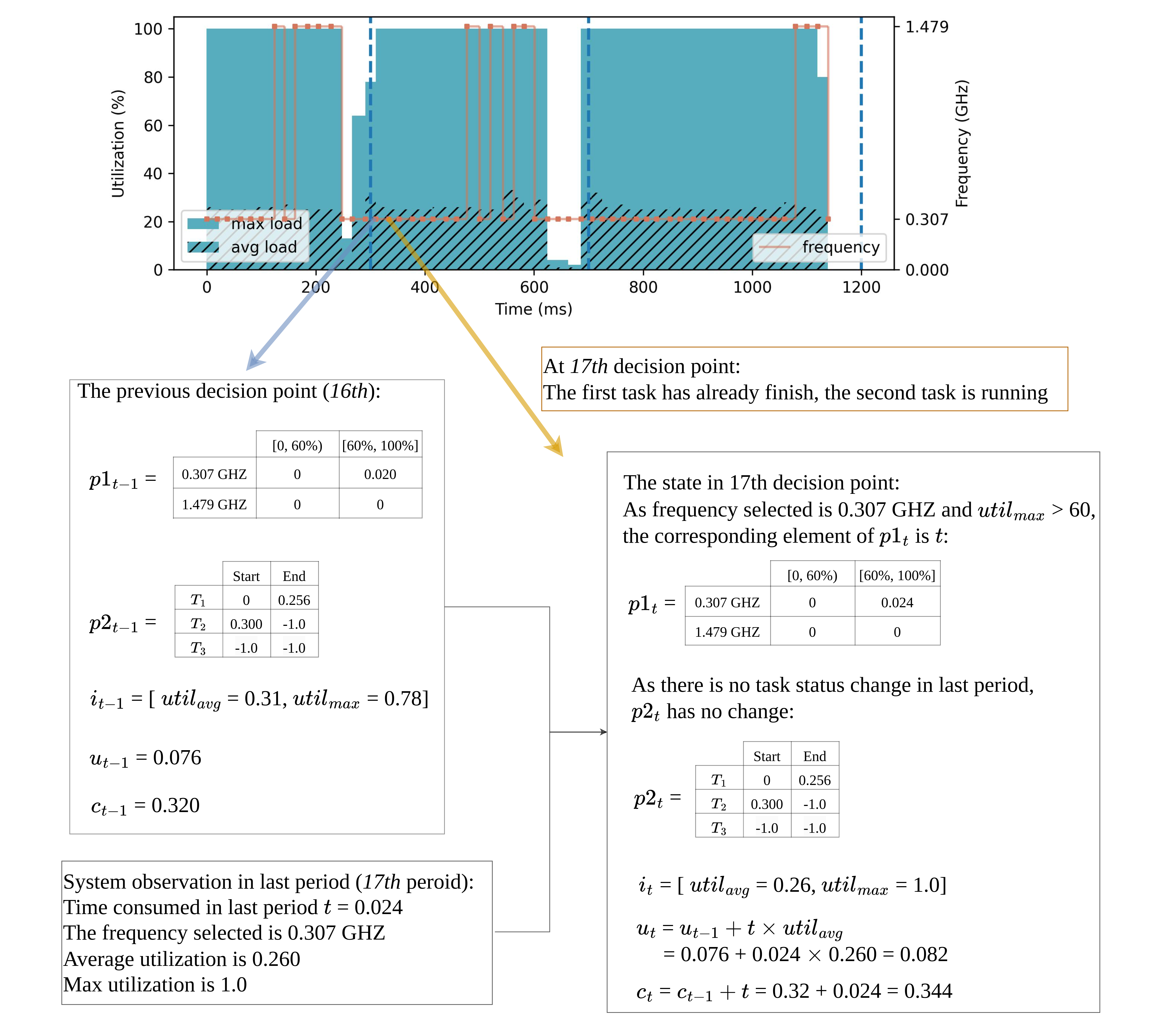}
  \caption{TEGM example.}
  \label{fig:exgru}
\end{figure}

The general calculating method is shown in Algorithm~\ref{alg:encoder}. 
Fig.~\ref{fig:exgru} shows an example of using this encoder with real system data. The figure shows the runtime situation of a three-task task set. The three tasks are the same and start at 0, 0.3, 0.7 second, the corresponding deadlines are 0.3, 0.4, 0.5 seconds. The vertical dash lines in the figure show the deadlines. The orange broken line is the frequency and the point on the line is the time that the governor makes the decision. The lateral distance between points is a period of time. The blue block is the max utilization while the slash is the average utilization for that period of time. This figure is the data that the governor makes the 17th decision. 
\textcolor{black}{The information under the blue arrow represents the information at the 16th decision point which is the input needed to make the 17th decision, corresponding to Line 1-2 of Algorithm~\ref{alg:encoder}. The first block under the blue arrow gives the detailed data of each part in $s_{t-1}$ (Line 2). The second block gives the last observation which is $freq$, $util_{max}$, $util_{avg}$ and $t$ (Line 1). This information, combined with the label of each task (Line 1), shown in the block beside the orange arrow, is needed to calculate $s_t$ (Line 3). In this case, the label of $T_1$ is 2 as it is completed, the label of $T_2$ is 1 as it is in progress and the label of $T_3$ is 0 as it has not started. There are two other sets of labels $start\_label_{i}$ and $end\_label_{i}$ for marking whether the task has started or finished. If the task has started or finished, the corresponding label will be 1, otherwise will be 0. This is used to record whether the task status has changed in the last period time or not. Line 4 is to calculate $p1_{t}$ which is a two-dimensional array based on $freq$ and $util_{max}$. The system will check the observation, then set the corresponding element of $p1_{t}$ to $t$. The next step is to calculate $p2_{t}$ corresponding to Line 5-11, which checks the label of each task. If the task has just started or finished in this period, then the corresponding element is set to $c_{t-1}$, which is the whole system-time consumed so far. Line 12-14 is to get $i_t$, $u_t$ and $c_t$ based on $t$, $util_{avg}$ and $util_{avg}$. A detailed example is shown in Fig.~\ref{fig:exgru}.}

\subsubsection{Temporal Encoder (TEM) for Network only using MLP}
\begin{figure}[h]
  \centering
  \includegraphics[width=\linewidth]{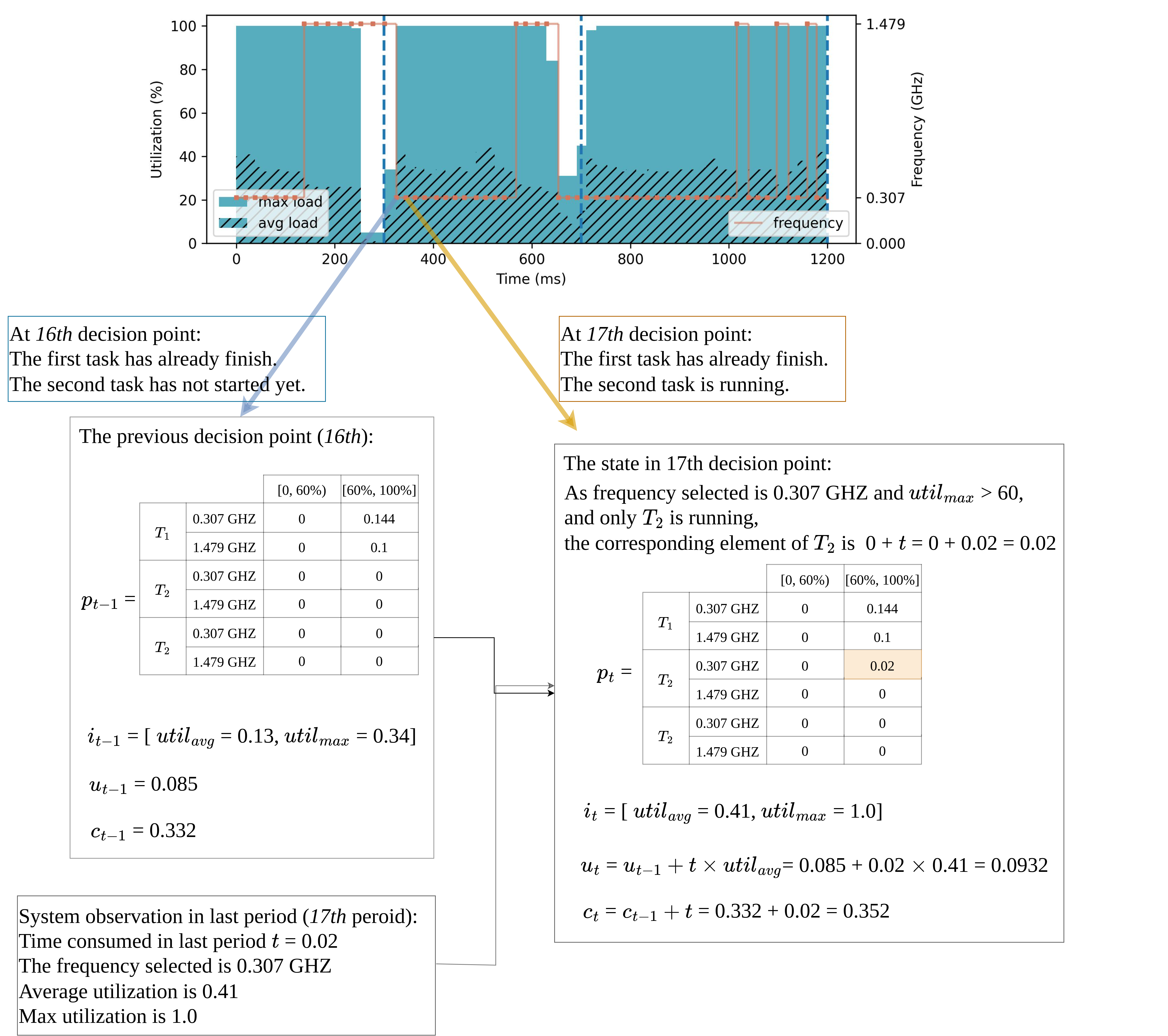}
  \caption{TEM example.}
  \label{fig:exmlp}
\end{figure}

The encoding method is shown in Algorithm~\ref{alg:mlpencoder}. 
\textcolor{black}{
Fig.~\ref{fig:exmlp} shows a detailed example of how the governor makes the 17th decision using this encoder.
Only the $p_t$ is different in this encoder (Line 4-8), all other parts are the same as TEGM Encoding (Line 9-12). $p_t$ in this encoding only has one part which is the accumulated execution time for each task, and dividing the time into multiple parts based on different frequencies and different loads. $p_t$ in TEM encoding is the same as $p1_t$ in TEGM encoding but has more details for each task.}

\begin{algorithm} 
    \caption{Temporal Encoder (\textcolor{black}{TEM}) For MLP} 
    \label{alg:mlpencoder} 
    \begin{algorithmic}[1]
    \State \textbf{INPUT:} The label $label_i$ sent from userspace for each task, the system observation of the current period $(freq, util_{max}, util_{avg})$ and time consumption for this period $t$
    \State $s_{t-1} = (p_{t-1}, i_{t-1}, u_{t-1}, c_{t-1})$
    \State \textbf{OUTPUT:} $s_t$
    \For{each label $label_i$}
        \If {$label_i$ is 1}
            \State $p_t[i][freq][load] = p_{t-1}[i][freq][load] + t$
        \EndIf
    \EndFor
    \State $i_t = (util_{max}, util_{avg})$
    \State $u_t = u_{t-1} + t \times util_{avg}$
    \State $c_t = c_{t-1} + t$
    \State $s_{t} = (p_{t},  i_{t}, u_{t}, c_{t})$
    \end{algorithmic}
\end{algorithm}

Tem coding method is more intuitive and simpler compared to TEGM coding, but less information is obtained. This encoding is suitable for small task sets.

\subsection{Inference Module in Kernel}
\label{sec:inferencemodule}
The policy is derived through the reinforcement learning method,  specifically the double deep Q learning. Like the previous work, the kernel only contains the inference part, which involves using the neural network to calculate the Q value for each candidate frequency. The governor will select the the frequency with maximum Q value for the next period of time.

The system running status is encoded as the state. As describe in Seciton~\ref{ch:encoding}, $s_t$ is consisted of $p_{t}, i_{t}, u_{t}, c_{t}$. The way to map the encoding to the frequency selected is to use neural networks. The two different encoding methods use different networks. Each time the governor makes the decision, $(s_t, freq_{next})$ is saved in the sequence $\phi$. If this sequence is used for training purposes, it will be sent to userspace for training.

For the first encoding method, a layer GRU is used to process $p1_t$. $p1_t$ is the data obtained at each system moment. Processing this time series data with GRU allows for the capture of temporal features well. The length of $p1_t$ is independent of the number of tasks in the task set, so the input to the GRU can have a fixed size. 

The data obtained through GRU processing, $pt\_fused$, is processed along with other data by a two-layer MLP. The candidate frequency is also fed into the MLP as an action to calculate the Q value. Then the governor selected based on the maximum Q value.

For the second method, the state and candidate actions are all processed through MLP. The problem is that the length sequence generated by this kind of encoding method is highly dependent on the number of tasks. The number of input nodes of MLP is proportional to the length of the \textcolor{black} {tasks}. Thus, it will be hard for the MLP to process a large scale of task sets.
When the number of tasks increases, the network will become larger. 

\begin{figure}[h]
  \centering
  \includegraphics[width=\linewidth]{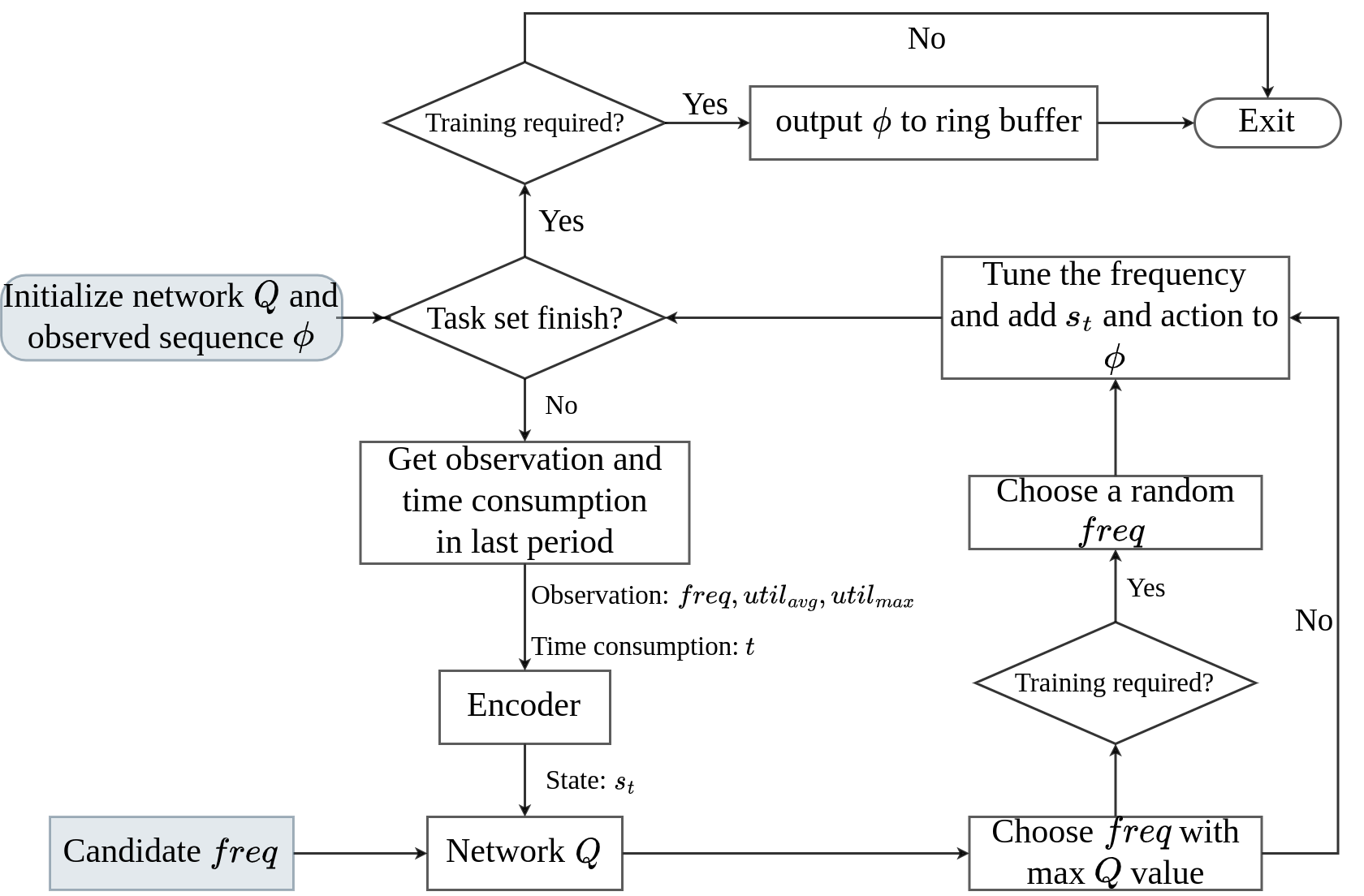}
  \caption{Kernel Inference Module.}
  \label{fig:kernel}
\end{figure}

The general inference way is similar to the previous work, shown in Fig~\ref{fig:kernel}. The difference between the two encoders is the network described before. When training is required, there is a certain probability that the frequency will be set to a random frequency for exploration. At the same time, the sequence will be recorded and sent to userspace for training.

% \begin{algorithm} 
%     \caption{Kernel Inference Module} 
%     \label{alg:kernelinfer} 
%     \begin{algorithmic}[1]
%     \State Initialize the network Q with parameters $\theta$ trained in userspace
%     \State Initialize the observed sequence $\phi$
%     \For{each sampling period}
%         \State get the observation form last period $(freq, util_{max}, util_{avg})$ and time consumption for this period $t$
%         \State Encode observed sequence to state $s_t$ based on temporal encoder
%         \State $freq_{next} = max_{freq}Q(s, freq; \theta)$
%         \If{training required}
%             \State Generate one random number $v$
%             \If{$v > \epsilon$}
%                 \State $freq_{next} = a random frequency$
%             \EndIf
%         \EndIf
%     \EndFor
%     \If{training required}
%         \State save $\phi$ in the ring buffer and output to userspace.
%     \EndIf

%     \end{algorithmic}
% \end{algorithm}

Note that in this work,  all the calculations use floating-point numbers. While there are security issues with using this within the kernel, the range is too wide for the data in this work to be normalized and quantified. Forcing the use of integers would instead create more of a burden. The codebase of building networks 
%\textcolor{black}{with quantitation?} 
in kernel can be found at: \url{https://github.com/coladog/tinyagent}.

\subsection{Training Module in Userspace}
The module gets the sequence $\phi$ from the kernel state. Based on the sequence, the module could calculate the reward following Algorithm~\ref{alg:reward}. The general calculation way is the same as the previous work, as shown in Algorithm~\ref{alg:reward}. It contains two components, the time under lower frequency and the average utilization. These two are for low energy consumption and high utilization. 
\textcolor{black}{Line 6 is to calculate the lower frequency part of the reward. For multiple frequencies that the system is in, the algorithm normalizes each frequency based on the cube of the maximum and minimum frequencies supported by the system and then inverts it. The time corresponding to each frequency is then multiplied. This way the more time the system is at a relatively low frequency, the higher the reward it will obtain. Line 7 is to get the utilization component for reward. Line 8 is to get the average of this frequency and utilization component.}

The difference is the penalty component, \textcolor{black}{corresponding to Line 9-18}. In previous work, once the task times out, the reward is set to 0, and the transitions after the timeout will be discarded. This work adds a threshold which denotes the acceptable exceed rate. Only when the actual exceed rate is larger than the setting threshold, the reward will be 0. 
\textcolor{black}{That is, when there are tasks that exceed the deadline, the algorithm calculates the percentage of execution of timeout tasks relative to their own deadline and takes the average, which is $e_r$ (Line 10-11). If this value exceeds the set threshold, then the reward is set to 0 (Line 13). If it does not, the original reward will be scaled down based on $e_r$ (Line 15).}
The penalty is set this way for the following reasons. 

\begin{itemize}
    \item In the case of multitasking, it may happen that the previous task times out while the following task does not. Therefore, directly discarding the part after the timeout may result in the discarding of useful training information. Therefore, only the timeout is detected in the reward calculation here.
    \item For one execution that there are tasks exceeding the deadline, a threshold is set to to give a penalty to those that are still within acceptable range. This is feasible for soft real-time systems. This penalty setting also ensures that the timeout reaward is not higher than the no-timeout case.
\end{itemize}

\begin{algorithm} 
    \caption{Reward Calculation} 
    \label{alg:reward} 
    \begin{algorithmic}[1]
    \State Let T denote the total time of one execution of task set
    \State Let $exet_i$ denote the execution time of $task_i$
    \State Let $ddl_i$ denote the deadline of $task_i$
    \State Let $n_e$ denote the number of tasks which exceed the deadline
    \State $x = \frac{time \; spend \; in \; f \; during \; T}{T} $
    \State $r_{freq} = \sum_{f \in F}(1-\frac{f^3-f^3_{min}}{f^3_{max}-f^3_{min}}) \times x$
    \State $r_{util} = average \; utilization \; during \; T$
    \State $r_t = \frac{f_{freq}}{2} + \frac{f_{util}}{2}$
    \If {there are tasks exceed the deadline}
        \State $e_r = \sum \frac{ddl_i}{exet_i}$
        \State $e_r = \frac{e_r}{n}$
        \If{$e_r$ is larger than threshold}
            \State $r_t = 0$
        \Else
            \State $r_t = \frac{r_t}{2} \times e_r \times 10$
        \EndIf
    \Else
        \State $r_t = r_t$
    \EndIf
    \end{algorithmic}
\end{algorithm}

The reward is sparse. Only the last transition in the episode is non-zero. The reason for doing so is that it is hard to judge whether the inter-state transition deserves a high reward or not. It is also hard to add a non-zero reward at every deadline because this information is not passed in this design. Therefore, at the end of the episode, the overall value of the episode is judged to be appropriate based on the overall timeout rate, energy consumption at low frequencies, and average utilization.

\begin{figure}[h]
  \centering
  \includegraphics[width=\linewidth]{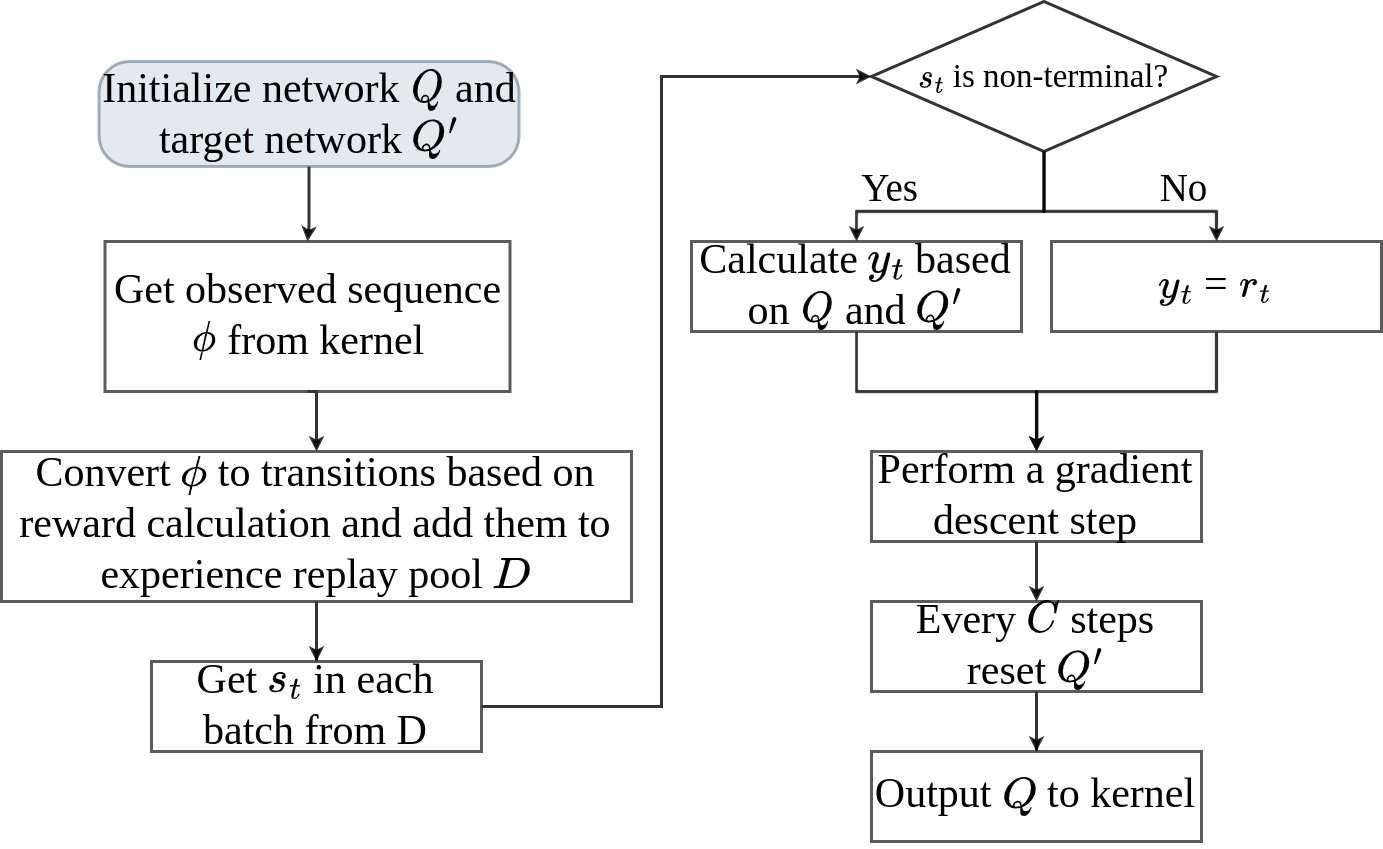}
  \caption{Userspace Training Module.}
  \label{fig:training}
\end{figure}

Training Module is shown in Fig.~\ref{fig:training}. It updates the parameters of the network based on DDQN. The updated data is sent to the kernel for the next round of training information gather. Based on this interaction, the kernel running generates the training data; the userspace performs training and updates the neural network.  After a certain amount of training, the governor could get a good policy.

% \begin{algorithm} 
%     \caption{Userspace Training Module} 
%     \label{alg:train} 
%     \begin{algorithmic}[1]
%     \State Initialize network Q with parameter $\theta$
%     \State Initialize target network Q' with parameter $\theta$' = $\theta$
%     \State Get observed sequence $\phi$ form kernel
%     \State Convert $\phi$ to transition based on the reward calculation and add each transition to experience relay pool D
%     \For{each batch in D}
%         \If{$s_t$ is non-terminal}
%             \State $y_t = r_t + \gamma Q'(s_{t+1}, maxQ(s_{t+1}, \alpha; \theta); \theta')$
%         \Else
%             \State $y_t = r_t$
%         \EndIf
%     \EndFor
%     \State Perform a gradient descent step
%     \State Every C steps reset $\theta$' = $\theta$.
%     \State Output $\theta$ into kernel space.
%     \end{algorithmic}
% \end{algorithm}

The experience replay pool follows the previous design. It is divided into 10 buckets. The range of reward belongs to [0, 1], which is divided into 10 levels and put into the bucket. Transitions with higher rewards have high priority.

%% file: Sections/experiment.tex
\section{Experiment}
The general architecture of the system in this work includes the kernel inference module and training module. These modules have been altered and optimized for multitasking scenarios.

\subsection{Experiment Setup}
The detailed experiment setup is shown in Table~\ref{tab:experiment}. The device used is the Nvidia Jetson Nano Board. We only allow two candidate frequencies which correspond to two-level voltages.
\textcolor{black}{It is worth mentioning that although our method is implemented on only one device and allows two F/V levels, the method itself is general and can be implemented for multiple F/V levels. The reason for choosing such a device is that our method is mainly aimed at small embedded devices.}
The network has two different settings. One network consists of a one-layer GRU and a two-layer MLP. The input and output of GRU are 2 and 6 respectively. The input of MLP depends on the task number. The number of two hidden layers is 30, and the number of outputs is 1. Another network only contains the MLP, the structure of the network is the same as the other. 
\textcolor{black}{The reason for setting 30 as the the number hidden node is to accommodate the varying input size. The input size of MLP is $GRU\,hidden\,size + task\,number \times 2 + 3$. The GRU hidden size is 6 which is fixed. If the task number is 3, then the input size of MLP is 15. Thus, the hidden size is set to twice the input size. However, if the task number is large, 30 hidden nodes are not enough. This will be explained in Section~\ref{sec:rewardcurve}.}
As mentioned in Section~\ref{ch:encoding}, we will use PRO. and PROM. for the abbreviation. In addition, we later expanded the first network to accommodate the increased number of tasks. We use PROL. as an abbreviation.

\begin{table}[htbp]
  \centering
  \caption{Experiment Setup}
  \label{tab:experiment}
    \begin{tabular}{|c||p{4.5cm}|}
        \hline
        Hardware & Nvidia Jetson Nano 2GB Board\\
        \hline
        OS & Linux 4.9 \\
        \hline
        Energy measurement & Ruideng UM25C USB power meter \\
        \hline
        NN structure & 2-6 GRU and 2-layer MLP, the number of inputs depending on task number, or pure 2-layer MLP\\
        \hline
    \end{tabular}
\end{table}

We pre-defined three workloads for testing our method. The detail information of these three workloads are shown in Table~\ref{tab:workload}. 
\textcolor{black}{These three task sets are designed to test the dynamic tuning ability of the method in different conditions, such as sequential and parallel tasks, small task sets and large task sets. Here we considered the eight-task task set as the large task set as our method targets the small embedded devices. Except for these three task sets, the method will be tested on random task sets which are shown in Section~\ref{sec:randomtask} to show the general effectiveness of the method. Also, there are no limitations to the task as long as it can run on the device.}

\begin{table}[htbp]
  \centering
  \caption{Workloads used}
  \label{tab:workload}
    \begin{tabular}{|c||p{4.5cm}|}
        \hline
        \textit{Three-task task set} & Self-constructed, no parallel tasks (if no timeout)\\
        \hline
        \textit{Five-task task set} & From Mibench, consisting of four benchmarks, \textit{jpeg, qsort, fft, typeset}, has parallel tasks\\
        \hline
        \textit{Eight-task task set} & From Mibench, consisting of four benchmarks, \textit{jpeg, qsort, fft, typeset}, has parallel tasks \\
        \hline
    \end{tabular}
\end{table}

The \textit{three-task task set} is a simple self-constructed workload consisting of the same CPU-intensive benchmark. This benchmark is constantly doing multiplication operations which are executed sequentially. The start time of the three tasks are 0, 0.3, 0.7, and the deadlines are 0.3, 0.4, 0.5 seconds respectively. If no task times out, then the entire task set has no parallel running tasks and the total runtime is 1.2 seconds. This workload is constructed to test whether the algorithm can recognize different deadlines for the same task for the purpose of dynamic tuning.

The \textit{five-task task set} is used to test parallel tasks. The start time of five tasks are 0, 0, 0, 0, 1.0 and the deadlines are 0.3, 0.5, 0.8, 1.0, 0.3. The total execution time will be 1.3 seconds. These benchmarks are selected from Mibench \cite{guthaus2001mibench} and its implementation detail is hidden.

The \textit{eight-task task set} is similar to the \textit{five-task task set}. The start time of five tasks are 0, 0, 0, 0, 1.0, 1.0, 1.0, 1.0 and the deadlines are 0.3, 0.5, 0.8, 1.0, 0.3, 0.5, 0.8, 1.0. The total execution time is 2.0 seconds. This task set is used to test whether the small-scale network could handle a task set with a slightly larger number of tasks. For the above tasks, both coding methods were tested and compared with the Linux built-in governor. Our governor only has two candidate frequencies. For Linux built-in governor governors, the choice of frequencies is all the frequencies that are supported on the Jetson Nano Board.

\subsection{Reward Curve}
\label{sec:rewardcurve}

\begin{figure}[h]
  \centering
  \includegraphics[width=\linewidth]{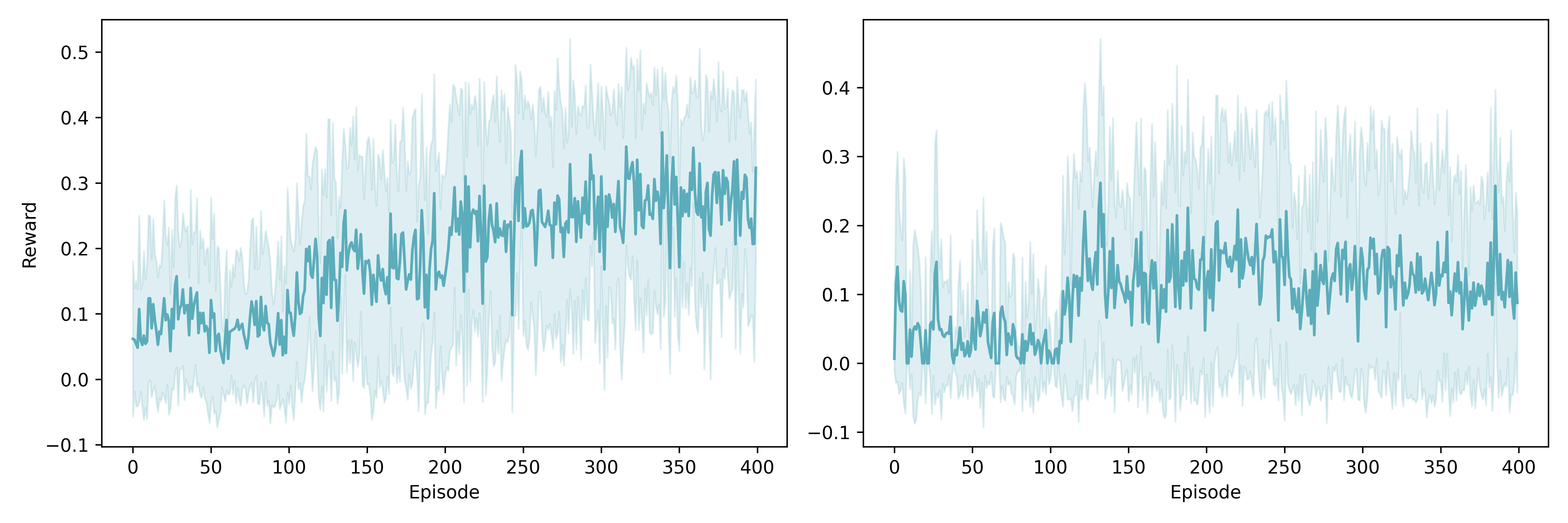}
  \caption{Reward curve with five training on \textit{three-task task set} with PRO. (left) and PROM. (right).}
  \label{fig:rewward3}
\end{figure}

\begin{figure}[h]
  \centering
  \includegraphics[width=\linewidth]{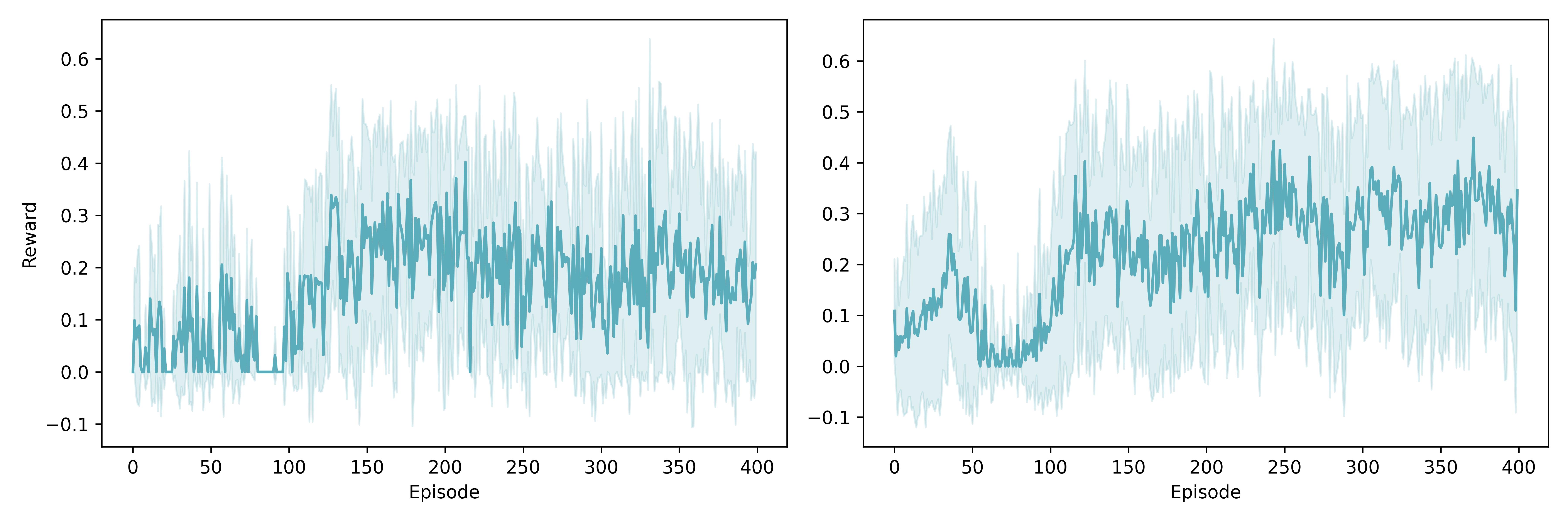}
  \caption{Reward curve with five training on \textit{Five-task taskset} with PRO. (left) and PROM. (right).}
  \label{fig:rewward5}
\end{figure}

After each training, we tested the updated network five times and recorded the average reward and its standard error. Fig.~\ref{fig:rewward3} and Fig.~\ref{fig:rewward5} show the reward cures for small-scale task sets (three and five) using two different encoders and networks. The overall reward curve shows an upward trend, showing that the reinforcement learning method acquires more rewards to generate better strategies. Especially for the \textit{three-task task set}, PRO. shows a steady upward trend which is better than PROM. While for the \textit{five-task task set}, these two show a similar trend. The PROM. is slightly better.

However, due to our penalty mechanism and the feature of the soft real-time system, a low reward does not necessarily mean that the generated strategy is bad. It could be that the governor is trying to exceed the deadline a little bit.

\begin{figure}[h]
  \centering
  \includegraphics[width=\linewidth]{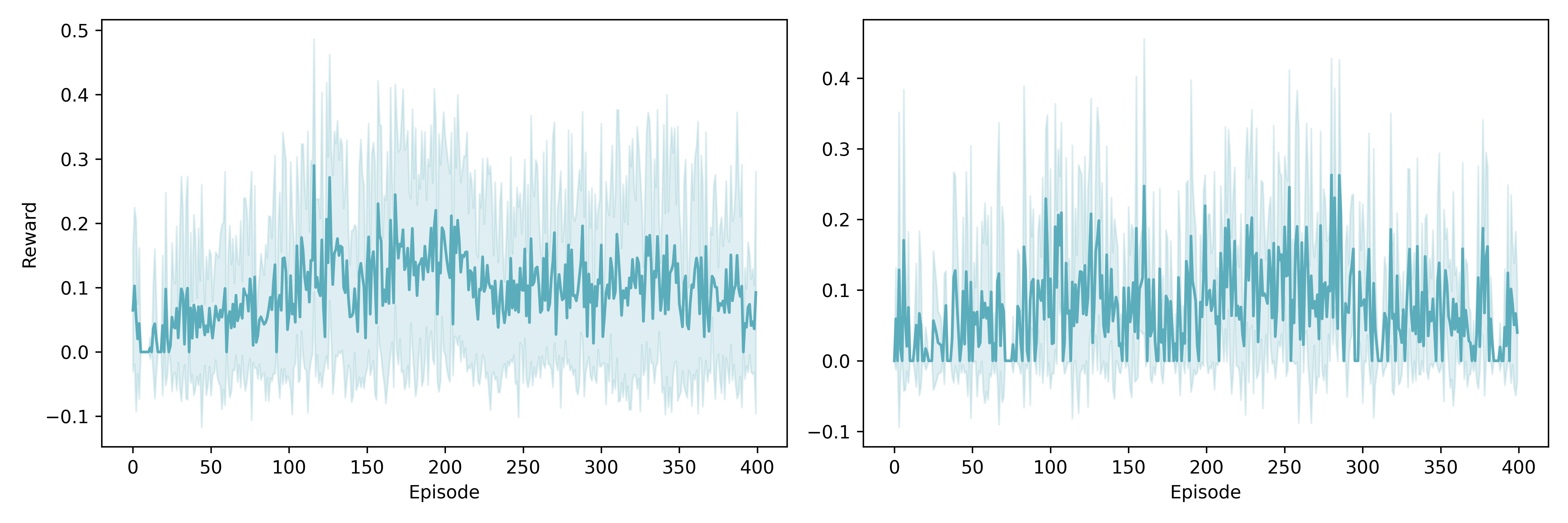}
  \caption{Reward curve with five training on \textit{eight-task task set} with PRO. (left) and PROL. (right).}
  \label{fig:reward8}
\end{figure}

We only used the PRO. and PROL. to test \textit{eight-task Taskset} since pure MLP networks are not suitable for large task sets. Fig.~\ref{fig:reward8} shows the reward curves, the left is a network with only 30 nodes in the middle layer, while the right is a network with 60 nodes. Looking only at the reward curve, increasing the number of nodes in the network does not lead to a better effect.

\subsection{Deadline Awareness}

\begin{figure}[h]
  \centering
  \includegraphics[width=\linewidth]{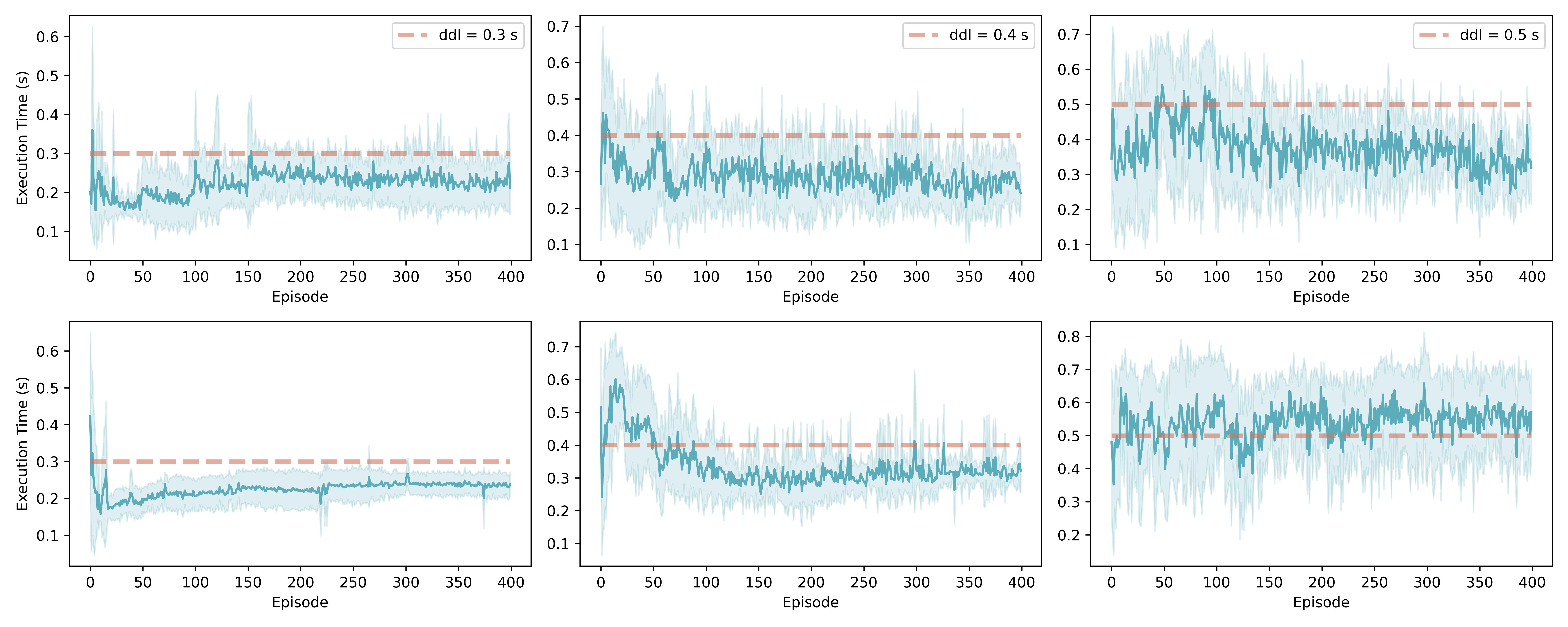}
  \caption{Execution time curve with five training on \textit{three-task task set} with  PRO. (top) and PROM. (bottom).}
  \label{fig:exet3}
\end{figure}

\begin{figure}[h]
  \centering
  \includegraphics[width=\linewidth]{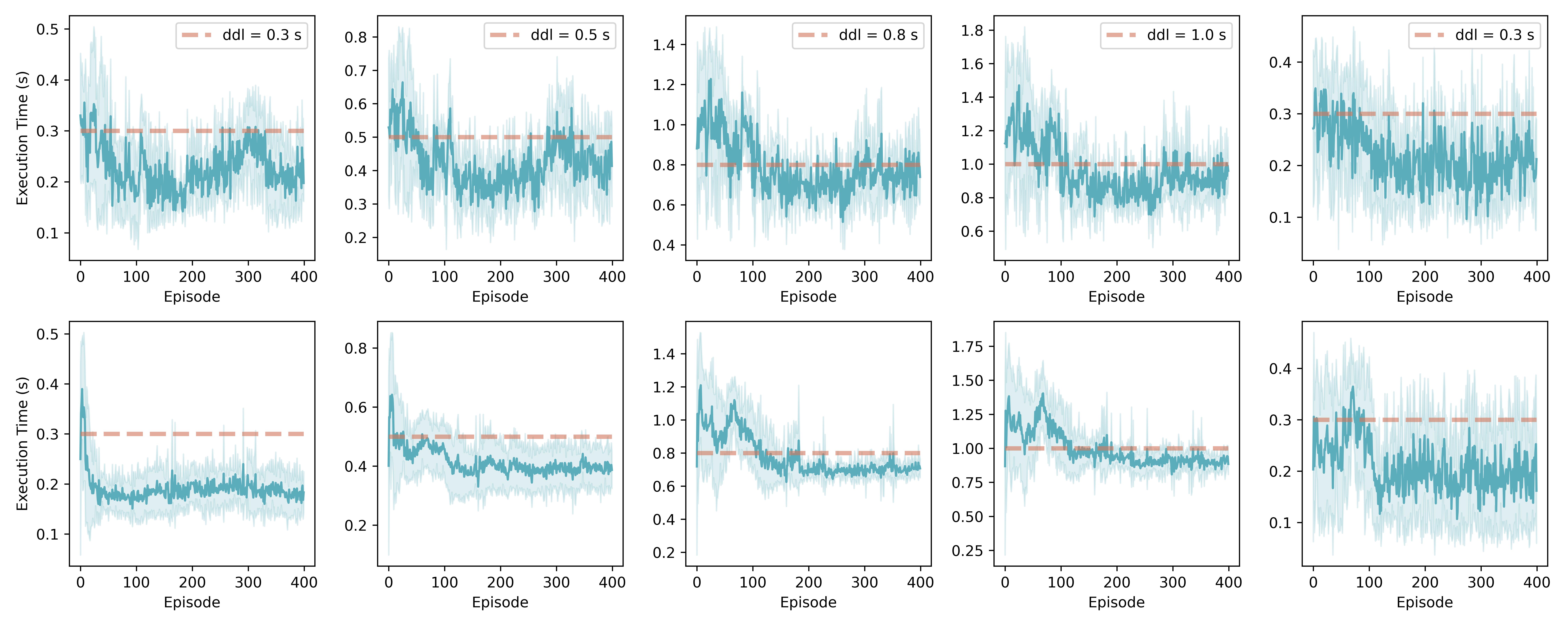}
  \caption{Execution time curve with five training on \textit{five-task task set} with PRO. (top) and PROM. (bottom).}
  \label{fig:exet5}
\end{figure}

\begin{figure}[h]
  \centering
  \includegraphics[width=\linewidth]{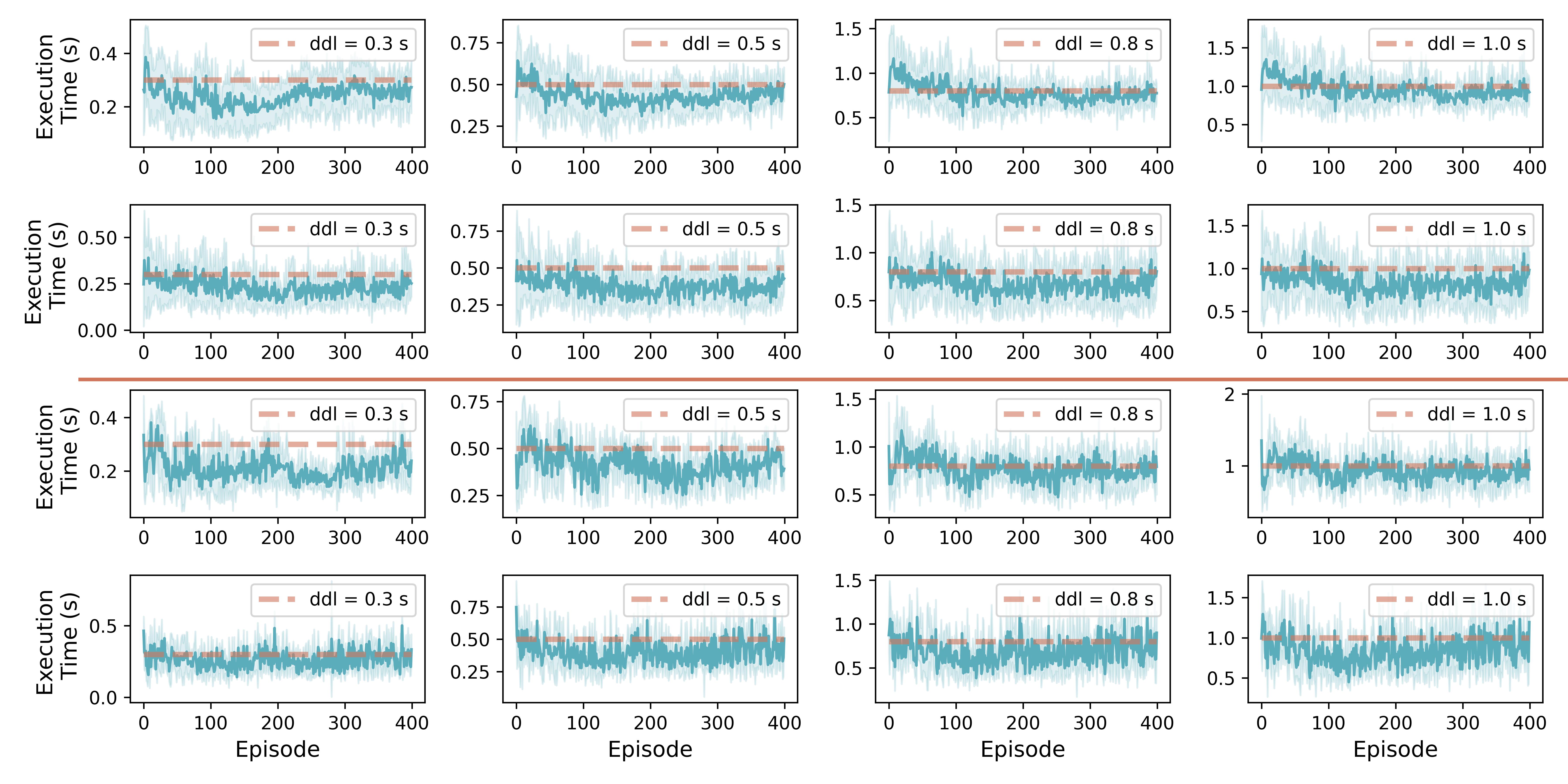}
  \caption{Execution time curve with five training on \textit{eight-task task set} with PRO. (first and second row) and PROL. (third and fourth row).}
  \label{fig:exet8}
\end{figure}

We also visualize the average execution time of each task after five tests. Fig.~\ref{fig:exet3}, Fig.~\ref{fig:exet5} and Fig.~\ref{fig:exet8} show the execution curve of three workloads respectively. The two parts divided by the orange horizontal line represent two encoders. In each part, one subfigure shows the execution time and deadline for one of the tasks in \textit{eight-task task set}.
For these three figures, each line represents a different network setting and each column represents the execution time of one task in the task set. The orange dashed line is the deadline for each task.

It can be seen from the figure that execution time is always near the deadline. Our proposed method could realize the different deadlines of different tasks or the same tasks very well, which is shown in Fig.~\ref{fig:exet3}. PRO. is better than PROM. since in the last task, the late execution time of PROM. is over the deadline.

For \textit{five-task task set}, PRO. performed better than PROM. since the execution time curve is closer to the deadline. Thus, comparing the two encoding methods, PRO. and PROM., PROM. is not as good as the RPO. with the increase of task number. For the \textit{eight-task task set} test, the larger network has a better result since the execution time curve is closer to the deadline in the latter four tasks.

\subsection{Learned Policy}
\textcolor{black}{
%Our proposed method could make different decisions for a similar load. The sequence of frequencies produced by our proposed method could be different for a similar load.In \textit{three-task task set}, the three tasks are the same but have different deadlines, both our methods could set the system to a low frequency at the beginning of the task to save energy and increase the frequency to meet the deadline. It could also control time at a low frequency to adapt to different deadlines. The two coding methods produce similar strategies with close results.
The sequence of decision made by the produced governor of our proposed method could be different for a similar load.
We observe that the produced strategy could set the system to a low frequency at the beginning of the task to save energy and increase the frequency to meet the deadline. The produced strategy could also control how long the system should be put at low frequency to adapt to different deadlines. The two coding methods produce similar strategies that can respect deadlines for three-task task set.}

% \begin{figure}[h]
%   \centering
%   \includegraphics[width=\linewidth]{Figure/policy_3.jpg}
%   \caption{Frequency policy developed by PRO. (top) and PROM. (bottom) on \textit{three-task task set}.}
%   \label{fig:policy3}
% \end{figure}

\begin{figure}[h]
  \centering
  \includegraphics[width=\linewidth]{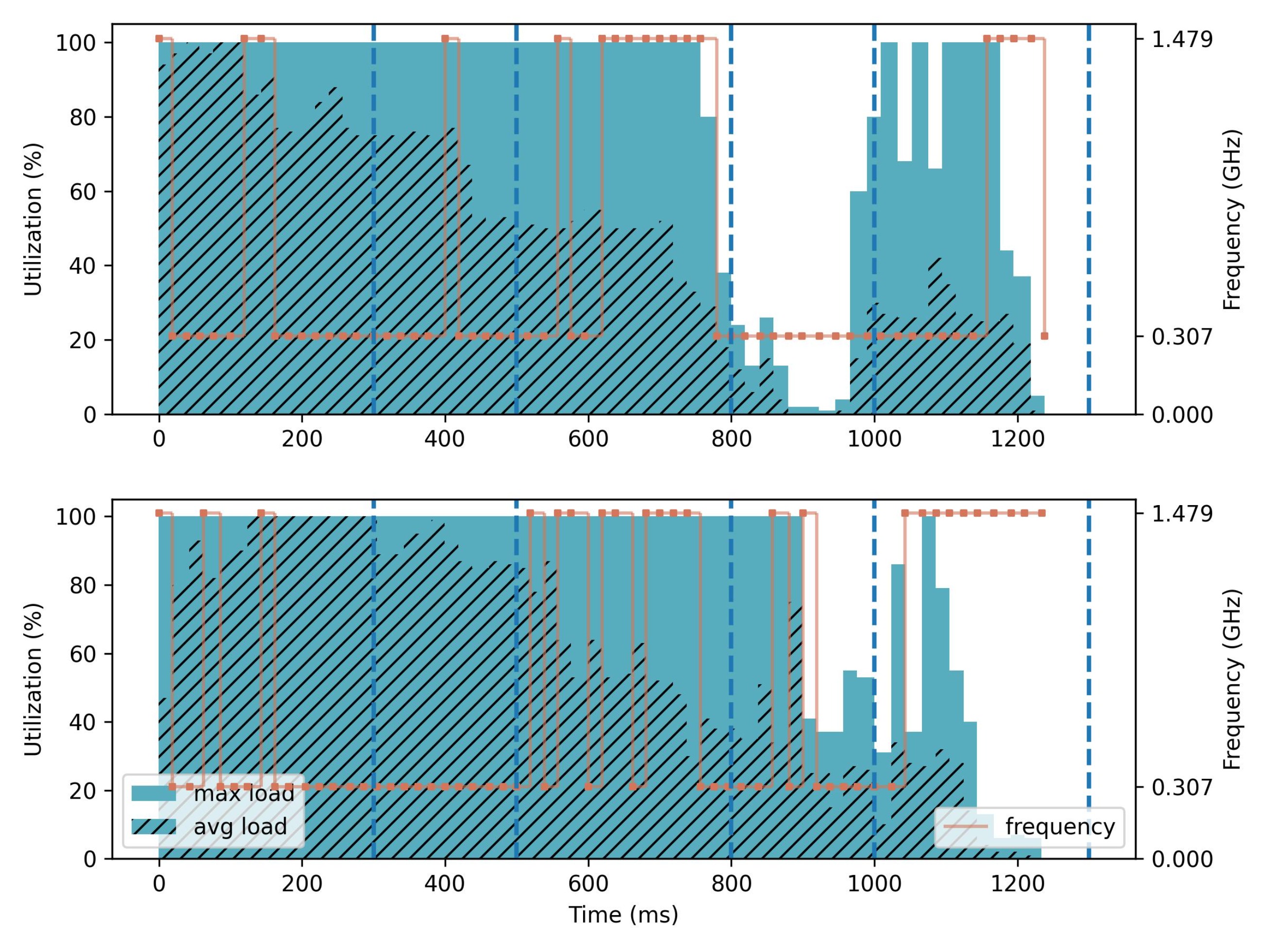}
  \caption{Frequency policy developed by PRO. (top) and PROM. (bottom) on \textit{five-task task set}.}
  \label{fig:policy5}
\end{figure}

\textcolor{black}{The same effect can be found on the policy for \textit{five-task task set}. We visualized the generated policy, as shown in Fig.~\ref{fig:policy5}. In the 200-400 ms phase, even though the average utilization of the system is almost 100\%, the governor still chooses a low frequency to reduce the energy consumption. While in the phase of 600-800 ms, the average utilization decreases, and the governor chooses a high frequency. For this task set, the PRO. performs better than the PROM.. When running the last task, the PRO. chose a low frequency at the beginning, while the PROM. chose a high frequency at the beginning and maintained this high frequency at the end. In this case, PRO. generated better policy.}

\textcolor{black}{For \textit{eight-task Taskset}, PROM. method did not get good results. PRO. can be applicable, but the number of middle-layer nodes needs to be expanded. Before expanding the number of nodes, PRO. cannot learn a good policy as its last task exceeds the deadline.}

% \begin{figure}[h]
%   \centering
%   \includegraphics[width=\linewidth]{Figure/policy_8.jpg}
%   \caption{Frequency policy developed by PRO. (top) and large GRU (bottom) on \textit{Eight-task Taskset}.}
%   \label{fig:policy8}
% \end{figure}

\subsection{Energy Saving}
\label{sec:energy}

\begin{figure}[h]
  \centering
  \includegraphics[width=\linewidth]{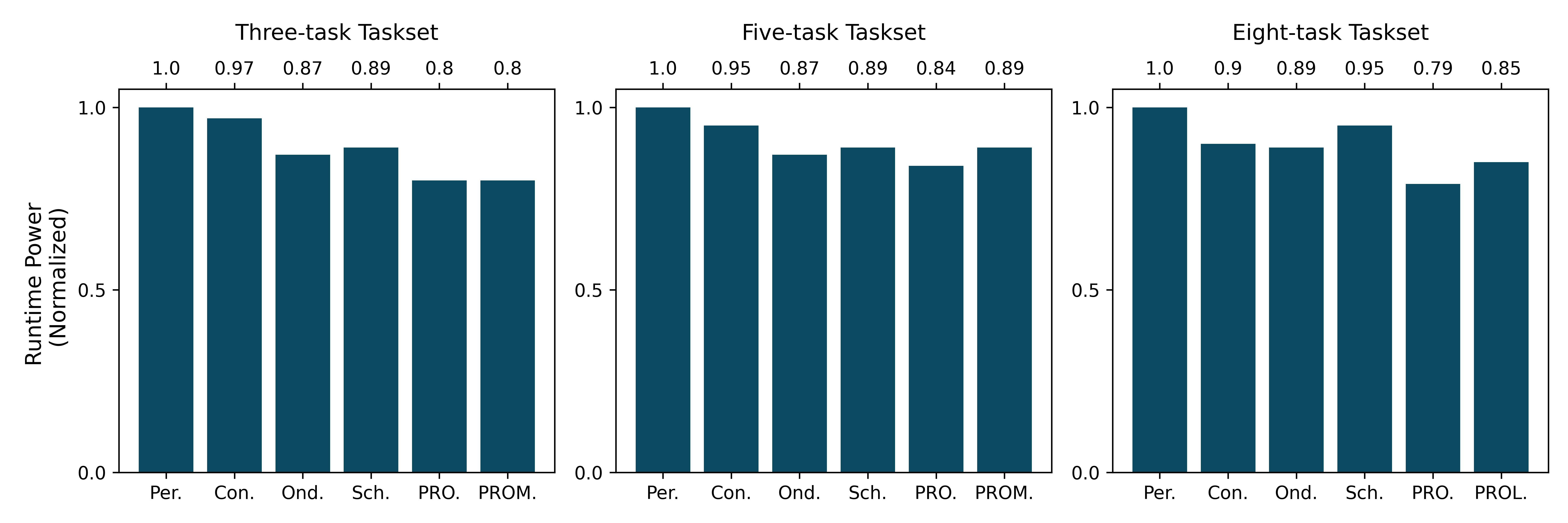}
  \caption{Power consumption on three tested task sets.}
  \label{fig:energy}
\end{figure}

Fig.~\ref{fig:energy} shows the normalized runtime power of three task sets. \textit{Per.}, \textit{Con.}, \textit{Ond.}, \textit{Sch.} represents four Linux built-in governors, \textit{Performance}, \textit{Conservative}, \textit{Ondemand} and \textit{Schedutil}. 
\textcolor{black}{Here we only considered these traditional Linux governors as baselines, as our method is also implemented Linux kernel. It is hard to compare with methods in other fields that are not implemented in the kernel.}
Table~\ref{tab:exet} shows the average execution time under different policies.

\begin{table}[htbp]
  \centering
  \caption{Execution time (s)}
  \label{tab:exet}
  \scalebox{0.8}{
    \begin{tabular}{|c|c|c|c|c|c|c|c|}
        \hline
        Taskset & Task & Per. & Con. & Ond. & Sch. & PRO. & PROM. \\
        \hline
        \multirow{3}*{\textit{Three-task Task set}} & task 1 & 0.138 & 0.200 & 0.150 & 0.171 & 0.266 & 0.262 \\
        \cline{2-8} 
        & task 2 & 0.139 & 0.187 & 0.159 & 0.185 & 0.344 & 0.331 \\
        \cline{2-8} 
        & task 3 & 0.139 & 0.184 & 0.158 & 0.189 & 0.345 & 0.447 \\
        \hline
        \multirow{5}*{\textit{Five-task Task set}} & task 1 & 0.100 & 0.444 & 0.100 & 0.132
        & 0.172 & 0.257 \\
        \cline{2-8} 
        & task 2 & 0.167 & 0.662 & 0.171 & 0.180 & 0.436 & 0.401 \\
        \cline{2-8} 
        & task 3 & 0.307 & 1.046 & 0.316 & 0.311 & 0.735 & 0.738 \\
        \cline{2-8} 
        & task 4 & 0.372 & 1.160 & 0.384 & 0.377 & 0.836 & 0.867 \\
        \cline{2-8} 
        & task 5 & 0.086 & 0.204 & 0.340 & 0.158 & 0.196 & 0.093 \\
        \hline
        \hline
        Taskset & Task & Per. & Con. & Ond. & Sch. & PRO. & PROL. \\
        \hline
        \multirow{8}*{\textit{Eight-task Task set}} & task 1 & 0.100 & 0.436* & 0.102 & 0.130
        & 0.223 & 0.093 \\
        \cline{2-8} 
        & task 2 & 0.167 & 0.663 & 0.171 & 0.181 & 0.307 & 0.163 \\
        \cline{2-8} 
        & task 3 & 0.304 & 1.072 & 0.312 & 0.316 & 0.745 & 0.759 \\
        \cline{2-8} 
        & task 4 & 0.371 & 1.208 & 0.384 & 0.379 & 0.938 & 0.841 \\
        \cline{2-8} 
        & task 5 & 0.095 & 0.276 & 0.284 & 0.131 & 0.131 & 0.092 \\
        \cline{2-8} 
        & task 6 & 0.167 & 0.431 & 0.362 & 0.182 & 0.267 & 0.377 \\
        \cline{2-8} 
        & task 7 & 0.303 & 0.702 & 0.485 & 0.318 & 0.817* & 0.521 \\
        \cline{2-8} 
        & task 8 & 0.371 & 0.797 & 0.510 & 0.375 & 1.135* & 0.789 \\
        \hline
    \end{tabular}
    }
\end{table}

We can see from the figure that PRO. and PROM. have similar results, both saving 7\% energy compared with \textit{Ondemand} for \textit{three-task task set}.  For \textit{Five-task task set}, both of our methods could learn a policy without exceeding the deadline. However, PROM. cannot save energy compared with \textit{Ondemand}. The PRO. method could save more energy (3\%). However, the original PRO. method could not learn a good policy for \textit{eight-task task set} since some of the tasks in the task set exceed the corresponding deadline too much, which can be seen in the table. After expanding the number of nodes, the governor could learn a good policy, which saves 4\% compared with \textit{Ondemand}.

\subsection{Inference Time Overhead}
% When the kernel uses the encoder and then infers through the network, there must be overhead compared with built-in governors. we tested the time required for these two parts with 8 tasks. The average overhead of the original PRO. is 0.09 ms. After expanding the nodes, the average overhead is 0.16 ms. The case of fewer tasks will have fewer overheads because the kernel does fewer floating-point calculations, as shown in Table~\ref{tab:inferencetime}.

\textcolor{black}{When the kernel uses the encoder and then infers through the network, there must be overhead compared with built-in governors. we tested the time required for these two parts with our tested task set. The inference time is based on the number of tasks as this decides the number of floating-point calculations made by the kernel. Also, it is affected by the policy our method generated. If the policy is inclined to choose high frequency, then the calculation time will be less, leading to a lower inference overhead, and vice versa. The results are shown in Table~\ref{tab:inferencetime}. The largest overhead is 0.16 ms. This is for \textit{eight-task task set} when using PROL. as it uses the largest network and has the most tasks. The second largest overhead is 0.12 ms for \textit{five-task task set} using PRO.. This is larger than the corresponding \textit{eight-task task set} as it use more lower frequency.
}

\begin{table}[htbp]
  \centering
  \caption{Inference Time (ms)}
  \label{tab:inferencetime}
  \scalebox{1}{
    \begin{tabular}{|c|c|c|c|}
        \hline
        Task Set & PRO. & PROM. & PROL.  \\
        \hline
        \textit{Three-task task set} & \textcolor{black}{0.08} &  \textcolor{black}{0.04} & / \\
        \hline
        \textit{Five-task task set} & \textcolor{black}{0.12} &  \textcolor{black}{0.06} & / \\
        \hline
        \textit{Eight-task task set} & 0.09 & / & 0.16 \\
        \hline
    \end{tabular}
    }
\end{table}
% 平均第一列推理时间: 109587.387283237 ns
% 平均第二列推理时间: 52330.20809248555 ns

% 平均第一列推理时间: 111083.4857142857 ns
% 平均第二列推理时间: 55057.86666666667 ns

% 平均第一列推理时间: 140303.09836065574 ns
% 平均第二列推理时间: 74353.32786885246 ns

% 平均第一列推理时间: 23772.612903225807 ns
% 平均第二列推理时间: 10312.572580645161 ns

% 平均第一列推理时间: 25419.860824742267 ns
% 平均第二列推理时间: 10653.953608247422 ns

% 平均第一列推理时间: 26851.74496644295 ns
% 平均第二列推理时间: 12921.778523489933 ns

\subsection{\textcolor{black}{Deadline missing}}
\textcolor{black}{As shown in Table~\ref{tab:exet}, we recorded the average execution time of each task in three task sets under different policies. Only the average execution time of Task 7 and Task 8 in \textit{Eight-task task set} under PRO. policy exceeds the deadline (with * mark in Table~\ref{tab:exet}). However, average execution time is in the deadline does not mean all the tasks in multiple tests are all in the deadline. Thus, we recorded and calculated the detailed percentage of tasks that did not miss the deadline and the extent of missing the deadline if the task exceeded the deadline. We only calculated the data for our proposed method and other baselines are not considered because traditional governors will not exceed the deadlines if the deadlines are reasonable. The policy of these baselines is stable, leading to a stable execution time.}

\textcolor{black}{The results are shown in Table~\ref{tab:deadlinemissing}. Here we only show the data for our policy, as other traditional governors will not exceed the deadline, except \textit{Conservative}. The policy used for \textit{Three-task task set} and \textit{Five-task task set} is PRO, and for \textit{Eight-task task set} is PROL. 0-2.5\% in the table means that $\frac{execution\_time - deadline}{deadline}$ belongs to (0, 2.5\%]. This is similar for 2.5-5\% and exceed 5\%. The data shows the percentage of all the tasks in one task set. For example, 87.17\% means that there are 87.17\% of Task 1, Task 2 and Task 3 in \textit{Three-task task set} that did not exceed the deadline. As shown in the table, for these three task sets, the majority of tasks (85\%) do not exceed the deadline. For those exceeding the deadline, there are only a small part of the tasks exceed 5\%. As our method targets on soft real-time system, we considered that under 5\% is acceptable. The results show that our policy has robustness in not missing deadlines, which ensures the performance constraint of the system.}

\begin{table}[htbp]
  \centering
  \caption{Average deadline missing}
  \label{tab:deadlinemissing}
  \scalebox{0.9}{
    \begin{tabular}{|c|c|c|c|c|}
        \hline
        Task Set & Not Exceed Deadline & 0-2.5\% & 2.5-5\% & Exceed 5\% \\
        \hline
        \textit{Three-task task set} & 87.17\% & 3.67\% & 4.33\% & 4.83\% \\
        \hline
        \textit{Five-task task set} & 85.74\% & 6.20\% & 4.10\% & 3.96\%  \\
        \hline
        \textit{Eight-task task set} & 87.18\% & 6.06\% & 4.25\% & 2.58\% \\
        \hline
    \end{tabular}
    }
\end{table}

\subsection{\textcolor{black}{Convergence and Training Time Analysis}}
\textcolor{black}{A certain epoch number is set for the training process. When our RL method reaches the specified epoch, it will stop. This way of training is not guaranteed to converge which means that it cannot completely ensure  learning an effective policy. However, in all our tests, the method learned a good policy both for our pre-defined fixed task sets and for randomly generated task sets with the epoch set to 400. Therefore, we consider this number of epochs to be sufficient for convergence for task sets with less than 8 tasks.}

\textcolor{black}{For the training time, our general training way is first collecting training data in the kernel, which is to run the task multiple times (3 times in our experiments). Then the data will be sent to userspace for training, and the new updated network will be sent back to the kernel. The time for this process is about 15 seconds for \textit{three-task task set}, 16 seconds for \textit{five-task task set} and 17 seconds for \textit{eight-task task set}. We repeated the process 400 times in our experiment, so for each task set, the total time to find a policy is around 1.667 hours, 1.778 hours and 1.889 hours.}

\subsection{\textcolor{black}{Random Task set Design and Test Results}}
\label{sec:randomtask}
\textcolor{black}{To validate the general ability of our proposed method to save energy and ensure performance when running different real tasks, we designed a random task generator. Then we used the task sets generated by the generator to test the method. The generator is shown in Algorithm~\ref{alg:randomtask}. The algorithm will output a task set $S$ with a total task number of $n$ based on a list of $B$ benchmarks where all are from the Mibench. Each benchmark $b_i$ from $B$ has its baseline time $bt_i$ which is used to generate its deadline. The baseline time is obtained by running the benchmark under \textit{Performance} governor. This means that the baseline time is the shortest execution time of the benchmark in the device. Line 8 of the algorithm is to get a $interval$ time based on randomly generated start time $start_i$ and $bt_i$. The deadline $ddl_i$ is picked from $interval$. }

\begin{algorithm}[h]
    \caption{\textcolor{black}{Random Task Generator}}
    \label{alg:randomtask} 
    \begin{algorithmic}[1]
        \State \textbf{Input:} List of benchmarks $B$ with their list of $BT$ and number of tasks $n$
        \State \textbf{Output:} A Task set $S$
        \State $S \gets \emptyset$
        \For{$i = 1$ to $n$}
            \State Randomly select a benchmark $b_i \in B$
            \State Get corresponding baseline time $bt_i$ based on $b_i$
            \State Randomly generate start time $start_i$
            \State $interval = (start_i + bt_i, \, start_i + 2 \times bt_i)$
            \State Randomly pick a $ddl_i$ from $interval$
            \State $S \gets S \cup \{(b_i, start_i, ddl_i)\}$
        \EndFor
        \State \textbf{return} $S$
    \end{algorithmic}
\end{algorithm}

\textcolor{black}{We generated four task sets with different task numbers, which are 3, 4, 5 and 8. The power consumption is compared between our proposed method and \textit{ondemand} policy. Here we only compared with \textit{ondemand} policy as this policy is the most energy-saving one for multitask with multiple deadlines task sets among traditional governors, which are already shown in previous experiment results (Fig.~\ref{fig:energy}). The results are shown in Table~\ref{tab:energy_savings}. The last column of the table shows the percentage of power consumption comparison between Ond. and PRO., which is calculated by $1 - \frac{power\,of\,PRO.}{power\,of\,Ond.}$. As can be seen from the results, our method is more energy efficient than \textit{ondemand}, with the best even reaching 15\%.}

\begin{table}[htbp]
  \centering
  \caption{Power Consumption (W) Comparison Between Ond. and PRO}
  \label{tab:energy_savings}
    \begin{tabular}{|c|c|c|c|}
        \hline
        Task Number & Ond. & PRO. & Saving (\%) \\
        \hline
        3 & 2.17700 & 1.84490 & 15.25 \\
        \hline
        4 & 2.21337 & 2.05280 & 7.25 \\
        \hline
        5 & 2.03960 & 1.87377 & 8.12 \\
        \hline
        8 & 1.97544 & 1.83226 & 7.26 \\
        \hline
    \end{tabular}
\end{table}

\textcolor{black}{The corresponding deadline missing results are shown in Table~\ref{tab:deadlinemissingrandom}. The percentage is similar to the fixed task sets which shows the effectiveness of our proposed method in random task sets.}

\begin{table}[htbp]
  \centering
  \caption{Average deadline missing for random task sets}
  \label{tab:deadlinemissingrandom}
  \scalebox{0.9}{
    \begin{tabular}{|c|c|c|c|c|}
        \hline
        Task Number & Not Exceed Deadline & 0-2.5\% & 2.5-5\% & Exceed 5\% \\
        \hline
        3 & 85.50\% & 2.33\% & 1.50\% & 10.67\% \\
        \hline
        4 & \textcolor{black}{83.72\%} & \textcolor{black}{1.03\%} & \textcolor{black}{2.68\%} & \textcolor{black}{12.57\%} \\
        \hline
        5 & 89.14\% & 4.20\% & 2.60\% & 4.06\%  \\
        \hline
        8 & 87.09\% & 3.88\% & 2.30\% & 4.59\% \\
        \hline
    \end{tabular}
    }
\end{table}

%% file: Sections/related_work.tex
\section{Related Work}

Small device computing \cite{minaeva2020control} \cite{nobrega2019iot} \cite{ramson2021self} %\cite{zheng2019energy}
\cite{zhou2023cpu} \cite{zhou2021deadline} %\cite{lee2022probabilistically} 
\cite{teslyuk2020optimal} \cite{manogaran2020isof} %\cite{salman2021systematic} 
%\cite{liu2020content} %\cite{yuan2021design} 
%\cite{chang2020real} 
%\cite{xu2024energy} 
\cite{wu2024intos} \cite{houssam2020hpc} \cite{chen2020scheduling} \cite{hoffmann2021online} \cite{wu2023devfuzz} \cite{wu2024intos} 
 \cite{zhan2019power} \cite{li2024deep} \cite{chen2021energy} \cite{fettes2018dynamic} \cite{panda2022energy} \cite{yeganeh2020ring} \cite{wang2021online} \cite{asghari2022combined} \cite{lin2023workload} has become a vital component of modern intelligent society and a prominent focus of recent research.
Distinct from experience-based heuristic designs, self-adaptive systems \cite{nobrega2019iot} \cite{ramson2021self}
\cite{zhou2023cpu} 
\cite{zhou2021deadline} \cite{teslyuk2020optimal} \cite{liu2020content} \cite{hoffmann2021online} 
\cite{zhang2023infinistore} \cite{ma2024malletrain} \cite{ali2025enabling}
\cite{li2024deep} \cite{fettes2018dynamic} \cite{panda2022energy} \cite{zhan2022deepmtl} \cite{zheng2019energy} 
\cite{yeganeh2020ring} \cite{wang2021online} \cite{asghari2022combined} \cite{han2024kace} \cite{wang2023general}
\textcolor{black}{\cite{xu2020self} \cite{golec2023atom}} have become a major area of interest for researchers.

\textcolor{black}{Among these self-adaptive systems, some works explore the advantage of combining RL and energy-saving methods such as DVFS, to build energy optimization systems.}
\textcolor{black}{Zheng and Louri \cite{zheng2019energy} combined DVFS, and power-gating (PG) to save both static and dynamic energy for Network-on-Chips (NoCs) where the architecture is a mesh topology consisting of routers. Each router has an RL agent that conducts Deep Q-learning. The RL method takes three voltage/frequency (V/F) levels as actions, and takes cache-related metrics, such as L1D cache miss, network-related metrics, such as local port	buffer utilization, and other parameters as states. The results show that they have a 26\% power reduction.
%The problem is this method only uses simulation to experiment instead of using real devices. 
%Yeganeh-Khaksar et al. \cite{yeganeh2020ring} saved energy for multicore real-time embedded systems using RL and DVFS. Compared with our method, this work is for sporadic tasks instead of periodic tasks. They use six V/S levels as actions. The state consists of per-task power consumption and transient faults. The core type ARM Cortex-A7 but this is a simulation and assumes it supports per-core DVFS which is quite hard to implement in the real environment.
Wang et al. \cite{wang2021online} considered both core and uncore frequency by using Double Q learning to put the uncore frequency at a low level to save energy while the core frequency at a high level to perform tasks. The RL states only use the instruction per cycle (IPC) and the misses per operation (MPO). The method is also assisted by the power capping technique, and the final method could save 12\% power with only a 3\% performance reduction.
% Asghar et al. \cite{asghari2022combined} used deep Q-network with coral reefs optimization and DVFS for cloud data centers. Coral reefs optimization is used for generating various solutions for task scheduling and deep Q-network is used for judging these solutions. This method still uses simulation.
Lin et al. \cite{lin2023workload} proposed an improved DVFS method for mobile devices considering both kernel level (CPU and GPU frequency) and hardware level (temperature). They use Deep Q Learning and its innovation is using a representation based on probability distribution as the state.
Li et al. \cite{li2024fidrl} considered time series data in the system including power consumption from the last state, temperature, the number of tasks completed and other parameters. The method uses deep Q learning to adjust V/F level and schedule tasks for the embedded system which reaches 23\% energy reduction.} 

\textcolor{black}{%All these works build the foundations for our work.
Compared with these works, our work considers a more recourse-limited real-time system with time constraints under the scenario of multitask and multiple deadlines. Also, our work is implemented on a real small device to validate the effectiveness of the method.}

%% file: Sections/conclusion.tex
\section{Conclusion}
In this work, we propose a DVFS method based on deep reinforcement learning to solve the problems that may occur in real-time systems with multiple tasks and multiple deadlines. Inserting a neural network within the governor in the kernel allows the system to adapt frequency control according to the current system and workload. This ensures both performance, that is no task exceeds its deadline, and energy savings. Two different encoders, TEM and TEGM, were tested for neural network state inputs. TEM achieves the same results as the TEGM for a small set of tasks, but after increasing the number of tasks, the TEGM is more effective and can train a better policy, testing with fixed sets of 3, 5 and 8 tasks, respectively. 
%to validate
%With the policy obtained using TEGM, for the three fixed task sets tested, the system could get 3\%-5\% energy savings over the traditional governor. 
For generalization, a generator that can generate random sets of tasks is used to generate multiple task sets. The test results for these task sets achieved similar results as the fixed task sets.

%% file: main.bbl
% Generated by IEEEtran.bst, version: 1.14 (2015/08/26)
\begin{thebibliography}{10}
\providecommand{\url}[1]{#1}
\csname url@samestyle\endcsname
\providecommand{\newblock}{\relax}
\providecommand{\bibinfo}[2]{#2}
\providecommand{\BIBentrySTDinterwordspacing}{\spaceskip=0pt\relax}
\providecommand{\BIBentryALTinterwordstretchfactor}{4}
\providecommand{\BIBentryALTinterwordspacing}{\spaceskip=\fontdimen2\font plus
\BIBentryALTinterwordstretchfactor\fontdimen3\font minus \fontdimen4\font\relax}
\providecommand{\BIBforeignlanguage}[2]{{%
\expandafter\ifx\csname l@#1\endcsname\relax
\typeout{** WARNING: IEEEtran.bst: No hyphenation pattern has been}%
\typeout{** loaded for the language `#1'. Using the pattern for}%
\typeout{** the default language instead.}%
\else
\language=\csname l@#1\endcsname
\fi
#2}}
\providecommand{\BIBdecl}{\relax}
\BIBdecl

\bibitem{minaeva2020control}
A.~Minaeva, D.~Roy, B.~Akesson, Z.~Hanz{\'a}lek, and S.~Chakraborty, ``Control performance optimization for application integration on automotive architectures,'' \emph{IEEE Transactions on Computers}, vol.~70, no.~7, pp. 1059--1073, 2020.

\bibitem{nobrega2019iot}
L.~N{\'o}brega, P.~Gon{\c{c}}alves, P.~Pedreiras, and J.~Pereira, ``An iot-based solution for intelligent farming,'' \emph{Sensors}, vol.~19, no.~3, p. 603, 2019.

\bibitem{ramson2021self}
S.~J. Ramson, W.~D. Le{\'o}n-Salas, Z.~Brecheisen, E.~J. Foster, C.~T. Johnston, D.~G. Schulze, T.~Filley, R.~Rahimi, M.~J. C.~V. Soto, J.~A.~L. Bolivar \emph{et~al.}, ``A self-powered, real-time, lorawan iot-based soil health monitoring system,'' \emph{IEEE Internet of Things Journal}, vol.~8, no.~11, pp. 9278--9293, 2021.

\bibitem{zhou2023cpu}
T.~Zhou and M.~Lin, ``Cpu frequency scheduling of real-time applications on embedded devices with temporal encoding-based deep reinforcement learning,'' \emph{Journal of Systems Architecture}, vol. 142, p. 102955, 2023.

\bibitem{lee2022probabilistically}
H.~Lee, Y.~Choi, T.~Han, and K.~Kim, ``Probabilistically guaranteeing end-to-end latencies in autonomous vehicle computing systems,'' \emph{IEEE Transactions on Computers}, vol.~71, no.~12, pp. 3361--3374, 2022.

\bibitem{teslyuk2020optimal}
V.~Teslyuk, A.~Kazarian, N.~Kryvinska, and I.~Tsmots, ``Optimal artificial neural network type selection method for usage in smart house systems,'' \emph{Sensors}, vol.~21, no.~1, p.~47, 2020.

\bibitem{manogaran2020isof}
G.~Manogaran, C.-H. Hsu, B.~S. Rawal, B.~Muthu, C.~X. Mavromoustakis, and G.~Mastorakis, ``Isof: information scheduling and optimization framework for improving the performance of agriculture systems aided by industry 4.0,'' \emph{IEEE Internet of Things Journal}, vol.~8, no.~5, pp. 3120--3129, 2020.

\bibitem{salman2021systematic}
S.~M. Salman, A.~V. Papadopoulos, S.~Mubeen, and T.~Nolte, ``A systematic methodology to migrate complex real-time software systems to multi-core platforms,'' \emph{Journal of Systems Architecture}, vol. 117, p. 102087, 2021.

\bibitem{liu2020continuous}
M.~Liu, X.~Ding, and W.~Du, ``Continuous, real-time object detection on mobile devices without offloading,'' in \emph{2020 IEEE 40th International Conference on Distributed Computing Systems (ICDCS)}.\hskip 1em plus 0.5em minus 0.4em\relax IEEE, 2020, pp. 976--986.

\bibitem{yuan2021design}
J.~Yuan, S.~Y. Samson, G.~Zhang, C.~P. Lim, H.~Trinh, and Y.~Zhang, ``Design and hil realization of an online adaptive dynamic programming approach for real-time economic operations of household energy systems,'' \emph{IEEE Transactions on Smart Grid}, vol.~13, no.~1, pp. 330--341, 2021.

\bibitem{wu2023energy}
Z.~Wu, L.~Han, J.~Liu, Y.~Robert, and F.~Vivien, ``Energy-aware mapping and scheduling strategies for real-time workflows under reliability constraints,'' \emph{Journal of Parallel and Distributed Computing}, vol. 176, pp. 1--16, 2023.

\bibitem{houssam2020hpc}
Z.~Houssam-Eddine, N.~Capodieci, R.~Cavicchioli, G.~Lipari, and M.~Bertogna, ``The hpc-dag task model for heterogeneous real-time systems,'' \emph{IEEE Transactions on Computers}, vol.~70, no.~10, pp. 1747--1761, 2020.

\bibitem{chen2020scheduling}
J.-J. Chen, J.~Shi, G.~von~der Br{\"u}ggen, and N.~Ueter, ``Scheduling of real-time tasks with multiple critical sections in multiprocessor systems,'' \emph{IEEE Transactions on Computers}, vol.~71, no.~1, pp. 146--160, 2020.

\bibitem{hoffmann2021online}
J.~L.~C. Hoffmann and A.~A. Fr{\"o}hlich, ``Online machine learning for energy-aware multicore real-time embedded systems,'' \emph{IEEE Transactions on Computers}, vol.~71, no.~2, pp. 493--505, 2021.

\bibitem{zhan2019power}
X.~Zhan, J.~Chen, E.~S{\'a}nchez-Sinencio, and P.~Li, ``Power management for multicore processors via heterogeneous voltage regulation and machine learning enabled adaptation,'' \emph{IEEE Transactions on Very Large Scale Integration (VLSI) Systems}, vol.~27, no.~11, pp. 2641--2654, 2019.

\bibitem{li2024deep}
X.~Li, L.~Chen, S.~Chen, F.~Jiang, C.~Li, W.~Zhang, and J.~Xu, ``Deep reinforcement learning-based power management for chiplet-based multicore systems,'' \emph{IEEE Transactions on Very Large Scale Integration (VLSI) Systems}, 2024.

\bibitem{chen2021energy}
X.~Chen, J.~Zhang, B.~Lin, Z.~Chen, K.~Wolter, and G.~Min, ``Energy-efficient offloading for dnn-based smart iot systems in cloud-edge environments,'' \emph{IEEE Transactions on Parallel and Distributed Systems}, vol.~33, no.~3, pp. 683--697, 2021.

\bibitem{fettes2018dynamic}
Q.~Fettes, M.~Clark, R.~Bunescu, A.~Karanth, and A.~Louri, ``Dynamic voltage and frequency scaling in nocs with supervised and reinforcement learning techniques,'' \emph{IEEE Transactions on Computers}, vol.~68, no.~3, pp. 375--389, 2018.

\bibitem{panda2022energy}
S.~K. Panda, M.~Lin, and T.~Zhou, ``Energy-efficient computation offloading with dvfs using deep reinforcement learning for time-critical iot applications in edge computing,'' \emph{IEEE Internet of Things Journal}, vol.~10, no.~8, pp. 6611--6621, 2022.

\bibitem{zhou2021deadline}
T.~Zhou and M.~Lin, ``Deadline-aware deep-recurrent-q-network governor for smart energy saving,'' \emph{IEEE Transactions on Network Science and Engineering}, vol.~9, no.~6, pp. 3886--3895, 2021.

\bibitem{zhou2023profiling}
T.~Zhou, H.~Wang, X.~Li, and M.~Lin, ``Profiling and understanding cpu power management in linux,'' in \emph{2023 IEEE Smart World Congress (SWC)}.\hskip 1em plus 0.5em minus 0.4em\relax IEEE, 2023, pp. 1--8.

\bibitem{van2016deep}
H.~Van~Hasselt, A.~Guez, and D.~Silver, ``Deep reinforcement learning with double q-learning,'' in \emph{Proceedings of the AAAI conference on artificial intelligence}, vol.~30, no.~1, 2016.

\bibitem{guthaus2001mibench}
M.~R. Guthaus, J.~S. Ringenberg, D.~Ernst, T.~M. Austin, T.~Mudge, and R.~B. Brown, ``Mibench: A free, commercially representative embedded benchmark suite,'' in \emph{Proceedings of the fourth annual IEEE international workshop on workload characterization. WWC-4 (Cat. No. 01EX538)}.\hskip 1em plus 0.5em minus 0.4em\relax IEEE, 2001, pp. 3--14.

\bibitem{wu2024intos}
Y.~Wu, B.~Min, M.~Ismail, W.~Xiong, C.~Jung, and D.~Lee, ``$\{$IntOS$\}$: Persistent embedded operating system and language support for multi-threaded intermittent computing,'' in \emph{18th USENIX Symposium on Operating Systems Design and Implementation (OSDI 24)}, 2024, pp. 425--443.

\bibitem{wu2023devfuzz}
Y.~Wu, T.~Zhang, C.~Jung, and D.~Lee, ``Devfuzz: automatic device model-guided device driver fuzzing,'' in \emph{2023 IEEE Symposium on Security and Privacy (SP)}.\hskip 1em plus 0.5em minus 0.4em\relax IEEE, 2023, pp. 3246--3261.

\bibitem{yeganeh2020ring}
A.~Yeganeh-Khaksar, M.~Ansari, S.~Safari, S.~Yari-Karin, and A.~Ejlali, ``Ring-dvfs: Reliability-aware reinforcement learning-based dvfs for real-time embedded systems,'' \emph{IEEE Embedded Systems Letters}, vol.~13, no.~3, pp. 146--149, 2020.

\bibitem{wang2021online}
Y.~Wang, W.~Zhang, M.~Hao, and Z.~Wang, ``Online power management for multi-cores: A reinforcement learning based approach,'' \emph{IEEE Transactions on Parallel and Distributed Systems}, vol.~33, no.~4, pp. 751--764, 2021.

\bibitem{asghari2022combined}
A.~Asghari and M.~K. Sohrabi, ``Combined use of coral reefs optimization and multi-agent deep q-network for energy-aware resource provisioning in cloud data centers using dvfs technique,'' \emph{Cluster Computing}, vol.~25, no.~1, pp. 119--140, 2022.

\bibitem{lin2023workload}
C.~Lin, K.~Wang, Z.~Li, and Y.~Pu, ``A workload-aware dvfs robust to concurrent tasks for mobile devices,'' in \emph{Proceedings of the 29th Annual International Conference on Mobile Computing and Networking}, 2023, pp. 1--16.

\bibitem{liu2020content}
Y.~Liu, H.~Jiang, Y.~Wang, K.~Zhou, Y.~Liu, and L.~Liu, ``Content sifting storage: Achieving fast read for large-scale image dataset analysis,'' in \emph{2020 57th ACM/IEEE design automation conference (DAC)}.\hskip 1em plus 0.5em minus 0.4em\relax IEEE, 2020, pp. 1--6.

\bibitem{zhang2023infinistore}
J.~Zhang, A.~Wang, X.~Ma, B.~Carver, N.~J. Newman, A.~Anwar, L.~Rupprecht, V.~Tarasov, D.~Skourtis, F.~Yan \emph{et~al.}, ``Infinistore: Elastic serverless cloud storage,'' \emph{Proceedings of the VLDB Endowment}, vol.~16, no.~7, pp. 1629--1642, 2023.

\bibitem{ma2024malletrain}
X.~Ma, F.~Yan, L.~Yang, I.~Foster, M.~E. Papka, Z.~Liu, and R.~Kettimuthu, ``Malletrain: Deep neural networks training on unfillable supercomputer nodes,'' in \emph{Proceedings of the 15th ACM/SPEC International Conference on Performance Engineering}, 2024, pp. 190--200.

\bibitem{ali2025enabling}
A.~Ali, X.~Ma, S.~Zawad, P.~Aditya, I.~E. Akkus, R.~Chen, L.~Yang, and F.~Yan, ``Enabling scalable and adaptive machine learning training via serverless computing on public cloud,'' \emph{Performance Evaluation}, vol. 167, p. 102451, 2025.

\bibitem{zhan2022deepmtl}
C.~Zhan, M.~Ghaderibaneh, P.~Sahu, and H.~Gupta, ``Deepmtl pro: Deep learning based multiple transmitter localization and power estimation,'' \emph{Pervasive and Mobile Computing}, vol.~82, p. 101582, 2022.

\bibitem{zheng2019energy}
H.~Zheng and A.~Louri, ``An energy-efficient network-on-chip design using reinforcement learning,'' in \emph{Proceedings of the 56th Annual Design Automation Conference 2019}, 2019, pp. 1--6.

\bibitem{han2024kace}
B.-S. Han, T.~Paul, Z.~Liu, and A.~Gandhi, ``Kace: Kernel-aware colocation for efficient gpu spatial sharing,'' in \emph{Proceedings of the 2024 ACM Symposium on Cloud Computing}, 2024, pp. 460--469.

\bibitem{wang2023general}
S.~Wang, R.~K. Williams, and H.~Zeng, ``A general and scalable method for optimizing real-time systems with continuous variables,'' in \emph{2023 IEEE 29th Real-Time and Embedded Technology and Applications Symposium (RTAS)}.\hskip 1em plus 0.5em minus 0.4em\relax IEEE, 2023, pp. 119--132.

\bibitem{xu2020self}
M.~Xu, A.~N. Toosi, and R.~Buyya, ``A self-adaptive approach for managing applications and harnessing renewable energy for sustainable cloud computing,'' \emph{IEEE Transactions on Sustainable Computing}, vol.~6, no.~4, pp. 544--558, 2020.

\bibitem{golec2023atom}
M.~Golec, S.~S. Gill, F.~Cuadrado, A.~K. Parlikad, M.~Xu, H.~Wu, and S.~Uhlig, ``Atom: Ai-powered sustainable resource management for serverless edge computing environments,'' \emph{IEEE Transactions on Sustainable Computing}, 2023.

\bibitem{li2024fidrl}
J.~Li, W.~Jiang, Y.~He, Q.~Yang, A.~Gao, Y.~Ha, E.~{\"O}zcan, R.~Bai, T.~Cui, and H.~Yu, ``Fidrl: Flexible invocation-based deep reinforcement learning for dvfs scheduling in embedded systems,'' \emph{IEEE Transactions on Computers}, 2024.

\end{thebibliography}
